\documentclass[aps,prd,showpacs,reprint,twoside,superscriptaddress,
               nofootinbib,showkeys,floatfix,twocolumn,a4paper]{revtex4-1}
\pdfoutput=1
\usepackage{amssymb,amsmath,amsfonts}
\usepackage{mathtools}
\usepackage{slashed}
\usepackage{braket}
\usepackage{graphicx}
\usepackage{hyperref}
\usepackage[html]{xcolor}
\usepackage{cleveref}
\usepackage{float}
\usepackage[utf8]{inputenc}
\graphicspath{{./Figs/}}
\newcommand{\Eq}[1]{Eq.~\eqref{#1}}
\newcommand{\Fig}[1]{Fig.~\ref{#1}}
\newcommand{\Tab}[1]{Table~\ref{#1}}

\newcommand{\MS}{\mathsf{\overline{MS}}}
\newcommand{\RI}{\mathsf{RI'}}
\newcommand{\RIMOM}{\mathsf{RI'\textnormal{-}MOM}}
\newcommand{\Tr}{\operatorname{Tr}}

\DeclareMathOperator{\rank}{rank}

\definecolor{mycol1}{HTML}{FF7F0E}
\definecolor{mycol2}{HTML}{1F77B4}
\definecolor{mycol3}{HTML}{2CA02C}
\definecolor{mycol4}{HTML}{D62728}
\definecolor{mycol5}{HTML}{9467BD}
\definecolor{mycol6}{HTML}{8C564B}
\definecolor{mycol7}{HTML}{E377C2}
\definecolor{mycol8}{HTML}{7F7F7F}
\definecolor{mycol9}{HTML}{BCBD22}
\definecolor{mycol10}{HTML}{17BECF}
\definecolor{mycol11}{HTML}{F9DC5C}
\newcommand\crule[3][black]{\textcolor{#1}{\rule{#2}{#3}}}

\begin{document}
\title{Nucleon generalized form factors from two-flavor lattice QCD}
\author{Gunnar Bali}
\email{gunnar.bali@ur.de}
\author{Sara Collins}
\email{sara.collins@ur.de}
\author{Meinulf G\"ockeler}
\email{meinulf.goeckeler@ur.de}
\author{Rudolf R\"odl}
\email{rudolf.roedl@ur.de}
\author{Andreas Sch\"afer}
\email{andreas.schaefer@ur.de}
\affiliation{Institut f\"ur Theoretische Physik, Universit\"at Regensburg,
             93040 Regensburg, Germany}
\author{Andr\'e Sternbeck}
\email{andre.sternbeck@uni-jena.de}
\affiliation{Theoretisch-Physikalisches Institut,
             Friedrich-Schiller-Universit\"at Jena, 07743 Jena, Germany}
\collaboration{RQCD collaboration}
\noaffiliation
\date{July 12, 2019}

\begin{abstract}
  We determine the generalized form factors, which
  correspond to the second Mellin moment (i.e., the first
  $x$-moment) of the generalized parton distributions of the nucleon
  at leading twist. The results are obtained using lattice
  QCD with $N_f=2$ nonperturbatively
  improved Wilson fermions, employing a range of quark masses down to an
  almost physical value with a pion mass of about 150\,MeV.
  We also present results for the isovector quark angular momentum and
  for the first $x$-moment of the transverse quark spin density.
  We compare two different fit strategies and find that directly fitting
  the ground state matrix elements to the functional form expected
  from Lorentz invariance and parametrized in terms of form factors
  yields comparable, and usually more stable results than the traditional
  approach where the form factors are determined from 
  an overdetermined linear system based on the fitted matrix elements.
\end{abstract}
\maketitle

\section{Introduction}
\label{sec:intro}
The understanding of hadron structure has greatly evolved over the last decades.
The collected knowledge is parametrized by a large number of functions.
Generalized parton distributions (GPDs) are one set of such functions.
They parametrize, e.g., the transverse coordinate distribution of partons
in a fast moving hadron and contain information on how these distributions
depend on the parton or hadron spin direction. Pinning down all these multivariable functions experimentally is unrealistic at present.
Therefore, lattice QCD has to substitute some of the missing experimental data.
With this article we contribute to the effort of various lattice groups to provide some of these needed results
\cite{Hagler:2007xi,Bratt:2010jn,Alexandrou:2011nr,Syritsyn:2011vk,Sternbeck:2012rw,Alexandrou:2013joa,Bali:2013dpa,Alexandrou:2013wka,Ji:2015qla,Bali:2016wqg,Chen:2016utp,Chen:2016fxx,Zhang:2017bzy}.

From the experimental point of view, GPDs play a similarly important role for the description of
exclusive hadronic reactions as parton distribution functions (PDFs) do for inclusive reactions. The most extensively studied channel
is deeply virtual Compton scattering (DVCS), i.e., Compton scattering with a
highly virtual incoming photon and a correspondingly large, spacelike
momentum transfer $Q^2=-q^2$.
One advantage of DVCS is that the GPD matrix element interferes with
the well-known Bethe-Heitler cross section for which the final state photon is emitted from
the scattered lepton. Thus the measured cross sections provide not only information on the
absolute value of the DVCS correlators but also on their signs.
In all generality, including spin effects, the experimental analysis
becomes somewhat involved, as is, e.g., illustrated by the publications~\cite{Airapetian:2001yk,Airapetian:2012pg} of the
\textsc{Hermes} experiment.
For a recent careful theoretical analysis and references to
experimental work see Ref.~\cite{Kumericki:2016ehc}.

The theoretical understanding of GPDs and their moments, the
generalized form factors (GFFs),
has already a long history and is presented in the seminal work
of Refs.~\cite{Dittes:1988xz,Mueller:1998fv,Ji:1996ek,Radyushkin:1996nd,Collins:1996fb}.
More recent reviews can be found in Refs.~\cite{Diehl:2003ny,Belitsky:2005qn}.
The interest in some of the nucleon GPDs (there exist in total eight) is increased by
the fact that they provide information on the elusive orbital angular momentum
of partons in the nucleon.
However, the physical interpretation in this case is not straightforward,
because there exist inequivalent definitions of orbital angular momentum~\cite{Jaffe:1989jz,Ji:1996ek}.
For recent discussions of this topic see, e.g., Refs.~\cite{Leader:2013jra,Ji:2015sio,Engelhardt:2017miy}
and the articles cited therein.
In this article we will not review the many fascinating aspects of GPDs but
concentrate on our lattice calculation of the nucleon GFFs
using well-established techniques for the calculation of Mellin moments of GPDs;
see, e.g., Ref.~\cite{Hagler:2009ni}.

We remark that recently new methods have been proposed
to obtain information on parton distribution functions
(PDFs), distribution amplitudes (DAs), transverse momentum dependent
PDFs (TMDPDFs) and GPDs that is complementary
  to the computation of Mellin moments with respect to Bjorken-$x$ from
  expectation values of local currents within external states, see, e.g.,
Refs.~\cite{Ji:2013dva,Lin:2014zya,Alexandrou:2015rja,Alexandrou:2016jqi,Bali:2018spj,Braun:2007wv}.
In these approaches Euclidean correlation functions are computed and
then matched within collinear factorization to light cone distribution
functions, employing continuum perturbative QCD.  For the example of
DAs~\cite{Bali:2018spj}, some of us are involved in calculations with these new
techniques, using the ``momentum smearing''
technique~\cite{Bali:2016lva} to enable large hadron momenta to be
realized, and found results that are consistent with, but less
accurate than those obtained from the lowest nontrivial
moment. This may change as smaller lattice
spacings and larger computers become available. Here we will only
determine the first $x$-moment, i.e., the second Mellin moment, to
constrain the nucleon GPDs.

This paper is organized as follows.
In Sec.\,\ref{sec:introGPD} we shortly review definitions
and the operator product expansion for Mellin moments of GPDs.
The lattice QCD techniques used to extract GFFs are introduced in
Sec.\,\ref{sec:extration_of_GFFs}
followed by a discussion of the numerical methods in Sec.\,\ref{sec:numerical methods}.
In Secs.\,\ref{sec:results_GFF} and~\ref{sec:chiral_J} we present our results.
Some preliminary findings have been reported in Refs.~\cite{Sternbeck:2012rw,Bali:2013dpa,Bali:2016wqg}.
Finally, we investigate the transverse spin density of the nucleon in
Sec.\,\ref{sec:nucleon_tomography}.

\section{Basic Properties of GPDs}
\label{sec:introGPD}
The starting point is the off-forward nucleon matrix element
\begin{equation}
\label{eq:bilocal-mat_1}
\mathcal{M}_q^\Gamma(x) =
  \int^\infty_{-\infty}\!\frac{\mathrm{d}\lambda}{4\pi}\,  e^{i\lambda x}
    \left\langle  N(p^\prime, \sigma^\prime)| O_q^\Gamma(\lambda)   |N(p,\sigma)  \right\rangle
\end{equation}
 of a bilocal operator with quark flavor $q$
\begin{equation}
 \label{eq:bilocal-mat_2}
    O_q^\Gamma(\lambda) = \bar{q}\left(-\lambda n/2\right) \,  \Gamma  \  \mathcal{U}_{-\lambda n/2}^{+\lambda n/2} \,  q \left(+\lambda n/2\right) \,.
\end{equation}
The Wilson line $\mathcal{U}$ in Eq.~(\ref{eq:bilocal-mat_2}) connects ${-\lambda n/2}$ and $+\lambda n/2$ on the light cone ($n^2=0$).
Depending on the Dirac structure, indicated by the symbol $\Gamma$ in Eqs.~(\ref{eq:bilocal-mat_1}) and  (\ref{eq:bilocal-mat_2}),
one can parametrize the matrix element $\mathcal{M}$ in terms of GPDs.
For leading twist these read (see, e.g., Refs.~\cite{Ji:1998pc,Hagler:2009ni}),
\begin{subequations}
\label{eq:gpds}
\begin{align}
\mathcal{M}_q^{\gamma^\mu}    = \
	&\overline{U}(p^\prime, \sigma^\prime)
		 \left[
		 \begin{pmatrix}
			\gamma^\mu\\
		\frac{ i\sigma^{\mu\nu}\Delta_\nu}{2m_N}
		 \end{pmatrix}
	\!\cdot\!
		 \begin{pmatrix}
			H^q\\
			E^q
		 \end{pmatrix}
		 \;\;
		 \right]
\!	U(p,\sigma) \, ,\\
\mathcal{M}_q^{\gamma^\mu\gamma_5}   = \
	&\overline{U}(p^\prime, \sigma^\prime)
		 \left[
		 \ \ \
		 \begin{pmatrix}
			\gamma^\mu \gamma_5\\
		 \frac{ \Delta^\mu \gamma_5  }{2m_N}
		 \end{pmatrix}
	\!\cdot\!
		 \begin{pmatrix}
			\widetilde{H}^q\\
			\widetilde{E}^q
		 \end{pmatrix} \, \;
		 \right]
\!	U(p,\sigma)\, , \\
\mathcal{M}_q^{i \sigma^{\mu\nu}} \!  = \
	&\overline{U}(p^\prime, \sigma^\prime)
		\left[
		 \begin{pmatrix}
		 i \sigma^{\mu\nu} \\
		 \frac{  \gamma^{[\mu} \Delta^{\nu]}  }{2m_N} \\
		 \frac{  \overline{p}^{[\mu} \Delta^{\nu]}  }{m_N^2} \\
		 \frac{  \gamma^{[\mu} \overline{p}^{\nu]}  }{m_N} \\
		 \end{pmatrix}
	\!\cdot\!
		 \begin{pmatrix}
			H_T^q\\
			E_T^q\\
			\widetilde{H}_T^q\\
			\widetilde{E}_T^q\\
		 \end{pmatrix}
		 \right]
\!	U(p,\sigma)\, ,
\end{align}
\end{subequations}
with $\sigma^{\mu\nu} = i\, [ \gamma^\mu , \gamma^\nu ]/2$
and the nucleon spinors
$\overline{U}(p^\prime, \sigma^\prime )$ and $U(p, \sigma)$.
The GPDs, e.g., $H^q$ and $E^q$, and the corresponding tensor structures
$\gamma^\mu$ and $i\sigma^{\mu\nu}\Delta_\nu/(2m_N)$
are written as vectors,
where we apply a standard scalar product to simplify the notation
and introduce the kinematic variables
\begin{align}
\Delta  \coloneqq p^\prime - p, \quad \quad
 \overline{p} \coloneqq (p^\prime + p)/2 \,.
\end{align}
For the antisymmetrization of indices we use the notation $[\ldots]$,
e.g., $B^{[\mu} C^{\nu]}\coloneqq B^{\mu}C^{\nu} - C^{\nu}
B^{\mu}\eqqcolon\mathsf{A}_{\mu\nu} B^{\mu}C^{\nu}$.
The GPDs are functions of the three variables $(x,\xi,\mathsf{t})$,
such that $H^q = H^q(x,\xi,\mathsf{t})$ etc.  We define
\begin{align}
\mathsf{t} \coloneqq \Delta^2 \leq 0, \quad  \quad
\xi        \coloneqq -\frac{n\cdot \Delta}{2},
\end{align}
where $\mathsf{t}$ is the total momentum transfer squared
which is related to the virtuality $Q^2 = -\mathsf{t}$.
The longitudinal momentum fraction $x$ varies between $-1$ and $1$ and
the skewness $\xi$ between $0$ and $1$. Negative values of
$x$ correspond to plus or minus (depending on the GPD)
times the corresponding antiquark GPD at $-x$.
In this work we restrict ourselves to the isovector case
and therefore we only consider the above eight quark GPDs.
An analogous set of gluonic GPDs exists, which we will not address here.
For a more detailed discussion we refer the reader to Refs.~\cite{Ji:1998pc,Vanderhaeghen:1999xj,Goeke:2001tz,Diehl:2003ny,Belitsky:2005qn,Hagler:2009ni}.

In physical terms (for $|x| > \xi$) GPDs parametrize the
probability amplitude for a hadron to stay intact if
a parton is removed at the light cone point $-\lambda/2$ and replaced by a
parton with different momentum at light cone time $\lambda/2$.
In practice, it is of crucial importance to find
effective parameterizations of GPDs with a minimum number of parameters
which are then fitted to experimental data see, e.g., Ref.~\cite{Kumericki:2016ehc}.
Lattice input in principle allows one to pin down the values of these
parameters; however, at present the accuracy of such studies is for many GPDs
not yet sufficient to make a decisive impact.

As time is analytically continued to imaginary time to enable the
numerical evaluation on the lattice, the light cone loses its meaning.
The operator product expansion (OPE) relates, however, Mellin moments of GPDs to
local matrix elements that are amenable to lattice calculation.
For $H^q$ and $E^q$, for instance, these $x$-moments read (see, e.g., Refs.~\cite{Ji:1998pc,Hagler:2009ni})
\begin{subequations}
\begin{align}
\int_{-1}^{+1}\!\!\mathrm{d}x \, x^{n-1} \, H^q(x, \xi, \mathsf{t})&=  \nonumber \\
\sum\limits_{i=0, \,\mathrm{even}}^{n-1}
(-2\xi)^i A^q_{ni}(\mathsf{t}) &+
(-2\xi)^{n} \,  C^q_{n0}(\mathsf{t})|_{n=\mathrm{even}}, \\
\int_{-1}^{+1}\!\!\mathrm{d}x \, x^{n-1} \, E^q(x, \xi, \mathsf{t})&=  \nonumber \\
\sum
    \limits_{i=0 ,\,\mathrm{even}}^{n-1}
	(-2\xi)^i B^q_{ni}(\mathsf{t})
	 &- (-2\xi)^{n} \,  C^q_{n0}(\mathsf{t})|_{n=\mathrm{even}}\,,
\end{align}
\end{subequations}
where the real functions
$A^q(\mathsf{t})$,
$B^q(\mathsf{t})$ and
$C^q(\mathsf{t})$ in the $\xi$-expansion on the
rhs are the GFFs.
The case $n=1$ corresponds to the electromagnetic form factors
$F_1^{q}(\mathsf{t}) = A^q_{10}(\mathsf{t})$
and
$F_2^{q}(\mathsf{t}) = B^q_{10}(\mathsf{t})$. For
$n=2$ and $\mathsf{t}=0$ we obtain
the average quark momentum fraction $A^q_{20} =  \langle x\rangle_{q^+}$,
where, for this example, we indicated $q^{\pm}=q\pm\bar{q}$. Below
we will drop this distinction since in the case of the vector
and tensor GPDs the even moments automatically give the $q^+$
combination and the odd moments $q^-$, while for axial GPDs it is
the opposite.

In principle one can determine Mellin moments of GPDs for any $n$ on the lattice,
in practice one is restricted to the lowest few $n$.
The reason for this restriction is twofold.
On the one hand the signal to noise ratio becomes worse for an increasing
number of covariant derivatives.
On the other hand as $n$ increases, mixing with lower-dimensional operators
will take place, resulting in divergences that are powers
of the inverse lattice spacing $a^{-1}$.
In this study we focus on the case $n=2$, where such mixing
does not occur.
Similarly to elastic form factors, the respective GFFs are extracted from lattice
calculations of two- and three-point correlation functions where the currents
are the local twist-2 operators,
 \begin{subequations}
 \label{eq:operators}
\begin{align}
    \mathcal{O}^{\mu \nu}_{V,q}(z) &= \mathsf{S}_{\mu\nu}\:\bar{q}(z) \,  \gamma^{\mu} i\overleftrightarrow{D}^{\nu}q(z)\,, \\
    \mathcal{O}^{\mu \nu}_{A,q}(z) &= \mathsf{S}_{\mu\nu}\:\bar{q}(z) \,  \gamma^{\mu}\gamma_5  i \overleftrightarrow{D}^{\nu}q(z)\,,\\
    \mathcal{O}^{\mu\nu\rho}_{T,q}(z) &= \mathsf{A}_{\mu\nu}\mathsf{S}_{\nu\rho}\:\bar{q}(z) i\sigma^{\mu\nu} i \overleftrightarrow{D}^{\rho}q(z)\,.
 \end{align}
\end{subequations}
 Here $\mathsf{S}_{\mu\nu}$ and $\mathsf{A}_{\mu\nu}$ denote symmetrization
 (also subtracting traces and dividing by $n!$ for $n$ indices) and
antisymmetrization operators, respectively, and
\begin{align}
\overleftrightarrow{D_\mu} \coloneqq \frac{1}{2}(\overrightarrow{D_\mu}-\overleftarrow{D_\mu})
\end{align}
is the symmetric covariant derivative.

In the continuum we can decompose the matrix elements
\begin{subequations}
\begin{align}
\label{eq:mat_master_v}
 \big\langle N(p^\prime,\sigma^\prime)|\mathcal{O}^{\mu\nu}_{V,q}|N(p,\sigma)\big\rangle \!  =  \! \overline{U}(p^\prime,\sigma^\prime)\mathbb{D}_{V,q}^{\mu \nu} U(p,\sigma)\, , \\
\label{eq:mat_master_a}
 \big\langle N(p^\prime,\sigma^\prime)|\mathcal{O}^{\mu\nu}_{A,q}|N(p,\sigma)\big\rangle \!  =  \! \overline{U}(p^\prime,\sigma^\prime)\mathbb{D}_{A,q}^{\mu \nu} U(p,\sigma)\, , \\
\label{eq:mat_master_t}
 \big\langle N(p^\prime,\sigma^\prime)| \mathcal{O}_{T,q}^{\mu\nu\rho}|N(p,\sigma)\big\rangle  \! =  \! \overline{U}(p^\prime,\sigma^\prime)\mathbb{D}_{T,q}^{\mu \nu \rho} U(p,\sigma) \, ,
\end{align}
\end{subequations}
with the nucleon four-momentum $(p^\mu) = (E_N(\vec{p}\,),\vec{p} \,)$.
In Sec.\,\ref{sec:extration_of_GFFs} we will show how we extract the matrix elements
from the temporal dependence of the three-point correlation functions.
The desired GFFs are contained in the Dirac structures,
\begin{subequations}
\label{eq:Decompositions}
\begin{align}
\label{eq:Decomposition_V}
\mathbb{D}^{\mu\nu}_{V,q}&=
\mathsf{S}_{\mu\nu}\,
\begin{pmatrix}
\gamma^{\mu}\overline{p}^{\nu} \\
i\sigma^{\mu\rho} \Delta_\rho \overline{p}^{\nu} /( 2 m_N)   \\
\Delta^{\mu}\Delta^{\nu} / m_N\\
\end{pmatrix}
\cdot
\begin{pmatrix}
A^q_{20}\\
B^q_{20}\\
C^q_{20}
\end{pmatrix} \,, \\
\label{eq:Decomposition_A}
\mathbb{D}^{\mu\nu}_{A,q} &=
\mathsf{S}_{\mu\nu}
\begin{pmatrix}
\gamma^{\mu}\gamma^{5}\overline{p}^{\nu}\\
\gamma_5  \Delta^\mu  \overline{p}^{\nu} / (2 m_N)
\end{pmatrix}
\cdot
\begin{pmatrix}
\widetilde{A}^q_{20} \\
\widetilde{B}^q_{20}
\end{pmatrix}  \,, \\
\label{eq:Decomposition_T}
\mathbb{D}^{\mu\nu\rho}_{T,q} &=
\mathsf{A}_{\mu\nu}
\mathsf{S}_{\nu\rho}\!
\begin{pmatrix}
    i \sigma^{\mu\nu} \overline{p}^{\rho}\\
    \gamma^{[\mu}\Delta^{\nu]}\overline{p}^{\rho} /(2 m_N)\\
    \overline p^{[\mu} \Delta^{\nu]}\overline{p}^{\rho} / m_N^2 \\
    \   \gamma^{[\mu}\overline{p}^{\nu]}\Delta^{\rho} / m_N
\end{pmatrix}
\cdot
\begin{pmatrix}
A^q_{T20}\\
B^q_{T20}\\
\widetilde{A}^q_{T20}\\
\widetilde{B}_{T21}
\end{pmatrix} \,.
\end{align}
\end{subequations}
Some aspects of GFFs have been more intensively discussed in the
literature than others, in particular,
\begin{itemize}
\item As has already been mentioned above,
  in the forward limit ($\mathsf{t}=0$), $A^q_{20}$ equals the average
  quark momentum fraction.
  Similar limits exist for
  $\widetilde{A}^q_{20}$ and $A^q_{T20}$ and the polarized and
  transversity PDFs, respectively.
\item Furthermore, in this limit $A^q_{20}$ and $B^q_{20}$ add up to twice
  the total angular momentum of the quark $q$ plus that of the antiquark
  $\bar{q}$ in the nucleon
   (the Ji sum rule~\cite{Ji:1996ek}) such that
  \begin{equation}
    J^{q} = \frac{1}{2}\left[A_{20}^{q}(0) + B^{q}_{20}(0)\right]
  \end{equation}
  represents the quark contribution to the nucleon spin.
  Combining $J^q$ with the quark spin contribution $\tfrac12\Delta\Sigma_q$,
  one can also obtain the quark orbital angular momentum
  $L_q = J_q-\tfrac12\Delta\Sigma_q$. We remark that this decomposition
  is not unique~\cite{Jaffe:1989jz}.
\item The five GFFs $A_{20}$, $B_{20}$, $A_{T20}$, $B_{T20}$ and
  $\widetilde{A}_{T20}$ parametrize, after Fourier transformation to
  impact parameter space, the first $x$-moment of the transverse spin
  density of a quark in a fast-moving nucleon~\cite{Burkardt:2000za}.
\end{itemize}

\section{Extracting generalized form factors}
\label{sec:extration_of_GFFs}
On the lattice, the GFFs are extracted from combinations of hadronic
two- and three-point correlation functions in Euclidean space-time.
The two-point function reads
\begin{align}
  \label{eq:2pt_cor}
  C^{2\mathrm{pt}}_{\alpha \beta}(t^\prime, \vec{p}^{\,\prime} ) =
  \sum_{\vec{x}^{\,\prime}} \! e^{-i\vec{p}^{\,\prime}\!\cdot\vec{x}^{\,\prime}}\left\langle \mathcal{N}_\alpha(t^\prime, \vec{x}^{\,\prime}) \, \overline{\mathcal{N}}_{\!\beta}(0, \vec{0}\,)\right\rangle,
\end{align}
where the nucleon destruction and creation interpolators $\mathcal{N}$ and $\overline{\mathcal{N}}$ are appropriate
combinations of $u$ and $d$ (anti)quark fields
\begin{subequations}
\label{eq:interplators}
\begin{align}
 \mathcal{N}_\alpha(t, \vec{x}\,) &= \varepsilon^{abc} u^a_{\alpha}(t, \vec{x}\,)\left[ u^b(t, \vec{x}\,)^{\intercal} \mathsf{C}\,\gamma_5 d^c(t, \vec{x}\,)\right], \\
 \overline{\mathcal{N}}_\beta(t, \vec{x}\,) &= \varepsilon^{abc} \left[ \bar{u}^b(t, \vec{x}\,)\, \mathsf{C}\gamma_5 \bar{d}^c(t, \vec{x}\,)^{\intercal}\right] \bar{u}^a_{\beta}(t, \vec{x}\,)\,.
\end{align}
\end{subequations}
$\mathsf{C}$ is the charge conjugation matrix.
The lattice three-point function is expressed as
\begin{align}
  \label{eq:3pt_cor}
  C^{3\mathrm{pt}}_{\alpha \beta}(\tau,t^{\prime}, \vec{p}^{\, \prime},\vec{p} \,) =
  \sum_{\vec{x}^{\, \prime} \vec{z}}
	  e^{-i\vec{p}^{\, \prime} \! \cdot \vec{x}^{\, \prime}}
  	  e^{+i\vec{z} \cdot ( \vec{p}^{\, \prime} -  \vec{p}\,)}  \nonumber \\
	      \times \left\langle  \!
	      \mathcal{N}_\alpha(t^{\prime}, \vec{x}^{\,\prime})
	      \mathcal{O}(\tau, \vec{z} \,)
	      \overline{\mathcal{N}}_{\!\beta}(0, \vec{0}\,)\right\rangle \,.
\end{align}
In this work we only consider isovector currents $\mathcal{O}$; therefore,
all quark lines are connected.
To improve the overlap of our interpolators
in Eqs.~\eqref{eq:interplators} with the physical ground state
we employ the combination of APE and Wuppertal (Gauss) smearing techniques
described in Refs.~\cite{Bali:2012av,Bali:2014nma,Bali:2014gha}.
This procedure reduces the impact of excited states substantially.
For the computation of \Eq{eq:3pt_cor},
we use the sequential propagator method~\cite{MAIANI1987420}
which implies fixing the sink time $t^\prime$.
We use the projector
\begin{align}
 \label{eq:projection}
\mathbb{P}^\rho=\frac{1}{2}\left(1+\gamma_4\right)\left(-i \gamma^\rho \gamma_5\right)^{1 + \delta_{\rho, 4}}
\end{align}
and contract it with the open spin indices of Eq.~\eqref{eq:3pt_cor}
to realize different spin projections and positive parity.
For $\rho=1,2,3$ we obtain the difference of
the spin polarization with respect to the quantization axis $\rho$, while
$\rho=4$ corresponds to the unpolarized case.
The positive parity projection is only correct for zero momentum;
however, excited state contributions (including states of different
parity for nonvanishing momentum) are exponentially suppressed at
large Euclidean times $\tau$. (The outgoing nucleon is projected onto
zero momentum.)

The definition of the operator $\mathcal{O}$ in \Eq{eq:3pt_cor}
depends on the desired GFF.
For the vector, axial and tensor GFFs at leading \mbox{twist-2}
the  operators are given in \Eq{eq:operators}.
On the lattice we construct our operators as linear combinations of
\begin{subequations}
	\label{eq:lat_ops}
\begin{align}
	\label{eq:lat_vec}
   \mathcal{O}^{\mu \nu}_{V,q}(z) &=
   \bar{q}(z)\gamma^{\mu} \overleftrightarrow{\nabla}^{\nu}q(z) \, , \\
   \mathcal{O}^{\mu \nu}_{A,q}(z) &=
   \bar{q}(z)\gamma^{\mu}\gamma_5 \overleftrightarrow{\nabla}^{\nu}q(z)  \, , \\
    \mathcal{O}^{\mu\nu\rho}_{T,q}(z) &= \bar{q}(z)i\sigma^{\mu\nu}\overleftrightarrow{\nabla}^{\rho}q(z) \, .
\end{align}
\end{subequations}
\begin{table}[tb]
  \caption{The renormalization factors used to translate our bare
    lattice data to the $\MS$ scheme at $\mu=2\,\text{GeV}$, obtained
    by reanalyzing the data of Ref.~\cite{Gockeler:2010yr}
      \label{tab:renfac}}
  \begin{center}
    \begin{ruledtabular}
      \begin{tabular}{c@{\qquad}c@{\quad}c@{\quad}c@{\quad}}
	& $\beta= 5.20$ & $\beta= 5.29$ & $\beta= 5.40$ \\
	\hline
	$Z_{\MS}^{v_{2,a}}$ \, & $1.090 \, (19)$ \, & $1.113 \,(15)$ \,  &  $1.140 \,(16)$ \\*[0.5ex]
	$Z_{\MS}^{v_{2,b}}$ \, & $1.096 \, (17)$ \, & $1.117 \,(21)$ \,  &  $1.143 \,(13)$ \\*[2ex]
	$Z_{\MS}^{r_{2,a}}$ \, & $1.083 \,(16)$ \, &  $1.106 \,(13)$ \, &  $1.134 \,(14)$ \\*[0.5ex]
	$Z_{\MS}^{r_{2,b}}$ \, & $1.118 \,(16)$ \, &  $1.138 \,(22)$ \, &  $1.163 \,(13)$ \\*[2ex]
	$Z_{\MS}^{h_{1,a}}$ \, & $1.115 \,(19)$ \, & $1.141 \,(19)$ \,  &  $1.171 \,(16)$ \\*[0.5ex]
	$Z_{\MS}^{h_{1,b}}$ \, & $1.129 \,(20)$ \, & $1.154 \,(20)$ \,  &  $1.184 \,(16)$
      \end{tabular}
    \end{ruledtabular}
  \end{center}
\end{table}

\begin{table}[tb]
  \caption{Relative error of the GFFs for the flavor
    combination $u-d$,
    induced by the uncertainty of the renormalization constants. This error
    turns out to be almost independent of the virtuality.
    \label{tab:pe}}
  \begin{center}
    \begin{ruledtabular}
      \begin{tabular}{ccccccccc}
        $A_{20}^{u-d}$ & $B_{20}^{u-d}$ & $\widetilde{A}_{20}^{u-d}$ & $\widetilde{B}_{20}^{u-d}$ & $A_{T20}^{u-d}$   & $B_{T20}^{u-d}$ & $\widetilde{A}_{T20}^{u-d}$ & $\widetilde{B}_{T21}^{u-d}$ &$\overline{B}_{T20}^{u-d}$ \\
        \hline
        0.019      & 0.019       & 0.015                  & 0.034                    & 0.020       & 0.020       & 0.020                  & 0.027                    & 0.020
      \end{tabular}
    \end{ruledtabular}
  \end{center}
\end{table}

In the case of the vector operator we work with multiplets that
transform according to two distinct irreducible representations
of the hypercubic group H(4) labeled as $v_{2,a}$ and $v_{2,b}$.
These are combinations of the operators in \Eq{eq:lat_vec} given by
\begin{align}
\label{eq:ov2a}
\mathcal{O}_{\mu \nu}^{v_{2,a}} &=  \mathsf{S}_{\mu\nu}\mathcal{O}_{\mu \nu}^V \quad \mathrm{with} \quad  1 \le \mu < \nu \le 4
\end{align}
and
\begin{subequations}
	\label{eq:ov2b}
	\begin{align}
	\mathcal{O}_{1}^{v_{2,b}}      &= \frac{1}{2}(\mathcal{O}_{11}^V + \mathcal{O}_{22}^V - \mathcal{O}_{33}^V - \mathcal{O}_{44}^V) \, ,\\
	\mathcal{O}_{2}^{v_{2,b}}      &= \frac{1}{\sqrt{2}}(\mathcal{O}_{33}^V-\mathcal{O}_{44}^V)\,  ,\\
	\mathcal{O}_{3}^{v_{2,b}}      &= \frac{1}{\sqrt{2}}(\mathcal{O}_{11}^V-\mathcal{O}_{22}^V) \,,
	\end{align}
\end{subequations}
respectively. The renormalized operators read
\begin{align}
\label{eq:z}
    \mathcal{O}_{\MS}^{v_{2,a\vert b}}(\mu) = Z(\beta, \mu)^{v_{2,a\vert b}}_{\MS} \, \mathcal{O}^{v_{2,a\vert b}}(\beta)\,,
\end{align}
where we use $\mu=2\,\text{GeV}$ as the renormalization scale.
Note that the renormalization factors depend on the multiplet, i.e.,
they slightly differ for $v_{2,a}$ and $v_{2,b}$.
Similarly, the axial operators are renormalized with
factors $Z_\MS^{r_{2,a}}$ and $Z_\MS^{r_{2,b}}$,
substituting $v_{2,a} \mapsto r_{2,a}$, $v_{2,b} \mapsto r_{2,b}$
in Eqs.~\eqref{eq:ov2a}, \eqref{eq:ov2b} and \eqref{eq:z}.
The tensor operators are renormalized with
$Z_\MS^{h_{1,a}}$ and $Z_\MS^{h_{1,b}}$.
The operator multiplets used in this case are listed in
Appendix~\ref{app:tensor_olc}.

A detailed description of the renormalization procedure, that consists
of first nonperturbatively matching from the lattice to the $\RIMOM$
scheme~\cite{Martinelli:1994ty,Chetyrkin:1999pq} and then
translating perturbatively to the $\MS$ scheme,
may be found in Ref.~\cite{Gockeler:2010yr}.
To make the article self-contained we summarize the basic steps
in Appendix~\ref{app:renormproc},
where we also address the error propagation from the renormalization
constants to the GFFs.
The relevant renormalization factors are summarized in \Tab{tab:renfac}.
They result from a reanalysis of the data presented
in Ref.~\cite{Gockeler:2010yr}
and correspond to the physical input $r_0=0.5\,\text{fm}$~\cite{Bali:2012qs}
and $r_0\Lambda^\MS=0.789$~\cite{Fritzsch:2012wq}.
\Tab{tab:pe} lists the relative errors on the renormalized
GFFs, associated with the uncertainties in the renormalization constants;
these amount to about~$2\%$.

\begin{table*}
  \caption{Parameters of the $N_f=2$ lattice ensembles used in this study.
    Latin numerals in the first column serve as ensemble identifiers.
    After the number of configurations $N_{\mathrm{conf}}$ we list in
    parentheses the number of independent (randomly chosen) source positions
    that we average over within each gauge configuration.
    Wherever this is indicated by parentheses after the sink-source
    separation $t'/a$, a smaller
    number of sources was used for this value. For more information about
    our setup we refer to Ref.~\cite{Bali:2014nma}.
    \label{tab:latsetup}}
  \begin{center}
    \begin{ruledtabular}
        \begin{tabular}{l@{\quad}c@{\quad}c@{\qquad}c@{\qquad}c@{\qquad}c@{\qquad}r@{\qquad}c@{\quad}c@{\quad} c@{\quad} }
        Ensemble             &
        $\beta$              & 
        $a$\,[fm]            &
        $\kappa$             &
        $V$                  &
        $m_{\pi}$\,[GeV]     &
        $m_N$    \,[GeV]     &
        $Lm_{\pi}$           &
        $N_{\mathrm{conf}}$  &
        $t'/a$\\ \hline
        \crule[mycol1]{0.3cm}{0.3cm} I&5.20&0.081&0.13596&$32^3\times 64$   &0.2795(18)& 1.091(08)   &3.69&$1986(4)$&13\\*[1.5ex]
        \crule[mycol2]{0.3cm}{0.3cm} II&5.29&0.071&0.13620&$24^3\times  48$ &0.4264(20)& 1.289(15)  &3.71&$1999(2)$&15\\
        \crule[mycol3]{0.3cm}{0.3cm} III\hfill&&&0.13620&$32^3\times 64$    &0.4222(13)& 1.247(06)           &4.90&$1998(2)$&15,17 \\
        \crule[mycol4]{0.3cm}{0.3cm} IV&&&0.13632 &$32^3\times 64$          &0.2946(14)& 1.071(11)      &3.42&$2023(2)$&7(1),9(1),11(1),13,15,17\\
        \crule[mycol5]{0.3cm}{0.3cm} V&&&&$40^3\times  64$                  &0.2888(11)& 1.079(09) &4.19&$2025(2)$&15\\
        \crule[mycol6]{0.3cm}{0.3cm} VI&&&&$64^3\times 64$                  &0.2895(07)& 1.072(05)    &6.71&$1232(2)$&15 \\
        \crule[mycol7]{0.3cm}{0.3cm} VII&&&0.13640&$48^3\times 64$          &0.1597(15)& 0.968(19)   &2.78&$3442(2)$&15 \\
        \crule[mycol8]{0.3cm}{0.3cm} VIII&&&&$64^3\times  64$&0.1497(13)    &0.944(17) &3.47&$1593(3)$&9(1),12(2),15\\*[1.5ex]
        \crule[mycol9]{0.3cm}{0.3cm} IX&5.40 & 0.060 &0.13640&$32^3\times 64$&0.490(02)&1.302(11) &4.81 & $1123(2)$ &   17 \\
        \crule[mycol10]{0.3cm}{0.3cm} X&&&0.13647&$32^3\times 64$&0.4262(20)&1.262(09) &4.18&$1999(2)$&17\\
        \crule[mycol11]{0.3cm}{0.3cm} XI&&&0.13660&$48^3\times 64$   & 0.2595(09) & 1.010(09) &  3.82   &   $2177(2)$   &    17
      \end{tabular}
      \end{ruledtabular}
  \end{center}
\end{table*}

In the following we demonstrate the extraction procedure
for the vector GFFs.
The axial and tensor GFFs are treated analogously.
We start by expanding \Eq{eq:3pt_cor} in terms of energy eigenstates
\begin{widetext}
\begin{align}
\label{eq:c3pt_a}
    C^{3\mathrm{pt}}_{\alpha \beta}(\tau,t^{\prime}, \vec{p}^{\, \prime},\vec{p}^{}\,)
=
	 \mathcal{A}_{\alpha \beta} \cdot
    \mathrm{e}^{-E_N(\vec{p}^{\, \prime}) \,(t^{\prime} - \tau)} \,
    \mathrm{e}^{-E_N(\vec{p}^{}\,)\,\tau} \,  + \, \mathrm{excited\,\,states}\,,
\end{align}
where the ground state amplitude reads
\begin{align}
\label{eq:c3pt_ab}
 \mathcal{A}_{\alpha \beta} =   \frac{1}{4 \, E_N(\vec{p}^{\, \prime}) E_N(\vec{p}^{}\,)}
 \sum \limits_{\sigma^\prime \sigma}
 \big\langle 0 | \mathcal{N}_\alpha | N(p^\prime,\sigma^\prime) \big\rangle
 \,\big\langle N(p^\prime,\sigma^\prime) |
    \mathcal{O}_{\MS}^{v_{2,a\vert b}}\,
    |
    N(p,\sigma) \big\rangle
  \,
 \big\langle N(p,\sigma)|\overline{\mathcal{N}}_{\!\beta}
  | 0 \big\rangle \,.
\end{align}
The exponentials contain the energy of the nucleon as a function of the considered spatial momentum,
the Euclidean operator insertion time $\tau$, and the sink time $t^{\prime}$.
Up to lattice artifacts, the matrix elements of an operator $\mathcal{O}^{\mu\nu}_{V,q}(z)$
can be decomposed according to the Euclidean versions of
Eqs.~\eqref{eq:mat_master_v} and \eqref{eq:Decomposition_V}.
In doing so, it is necessary to distinguish between the two
multiplets $v_{2,a}$
and $v_{2,b}$ [cf.\ Eqs.~\eqref{eq:ov2a} and \eqref{eq:ov2b}].
The decomposition can be written as
\begin{align}
 \left\langle N(p^\prime,\sigma^\prime) |
    \mathcal{O}_{\MS}^{v_{2,a\vert b}}\,
    |
    N(p,\sigma) \right\rangle =  \overline{U}(p^\prime,\sigma^\prime) \, \mathbb{D}_{\MS}^{v_{2,a\vert b}} \, U(p,\sigma)\,.
\end{align}
Applying the projection operator $\mathbb{P}^\rho$ [cf. \Eq{eq:projection}]
to $C^{3\mathrm{pt}}$ yields
\begin{align}
\label{eq:c3pt_ampl}
 & c^\rho_V(\tau,t^{\prime}, \vec{p}^{\, \prime},\vec{p}^{} \,) \coloneqq \sum_{\alpha,\beta} \mathbb{P}^\rho_{\beta \alpha} C^{3\mathrm{pt}}_{\alpha \beta}(\tau,t',\vec{p}^{\,\prime},\vec{p}^{}\,) =
 \sqrt{ Z(\vec{p}^{\,\prime}) \, Z(\vec{p}^{}\,)} \: \mathcal{F}_V \:\mathrm{e}^{-E_N(\vec{p}^{\, \prime}) \,(t^{\prime} - \tau)} \,
    \mathrm{e}^{-E_N(\vec{p}\,)\,\tau} \,  + \, \mathrm{excited\,\,states}
\end{align}
\end{widetext}
with
\begin{align}
\label{eq:ffunc}
 & \mathcal{F}_V =  \frac{
    \operatorname{tr} \left\{
       \mathbb{P}^\rho \,
       [ -i \slashed{p}^{\prime} + m_N] \:
\mathbb{D}_{\MS}^{v_{2,a\vert b}} \,
       [ -i \slashed{p} + m_N] \,
     \right\}
 }{4 \, E_N(\vec{p}^{\,\prime}) E_N(\vec{p}^{}\,)}
\end{align}
and
$
\slashed{p} \coloneqq i E_N(\vec{p}\,) \gamma_4 + \vec{p} \cdot \vec{\gamma} \,.
$
The $Z$ factors in \Eq{eq:c3pt_ampl} depend on the overlap of our nucleon interpolation operators with the nucleon ground state. They vary with momentum and smearing and can be extracted from the two-point correlation function $C^{2\mathrm{pt}}$.

The right-hand side of \Eq{eq:ffunc} contains the desired GFFs. The prefactors can be computed by inserting the respective Euclidean $\gamma$-matrices.
Here we restrict ourselves to the final momentum $\vec{p}^{\,\prime}=\vec{0}$.
Taking all available combinations of operators [cf.\ Eqs.~\eqref{eq:ov2a} and \eqref{eq:ov2b}],
projections $\mathbb{P}^\rho$ and momenta $\vec{p}$ for a fixed virtuality
\begin{equation}
\label{eq:virtuality}
Q^2=-\mathsf{t} = \left(\vec{p}^{ \, \prime}  - \vec{p}^{}\,\right)^2-
\left( \!\!\sqrt{m_N^2 + \vec{p}^{ \, \prime 2} } - \sqrt{m_N^2 + \vec{p}^{\, 2} } \right)^2 \,,
\end{equation}
we obtain a linear system of equations
\begin{align}
\label{eq:eqs_}
\vec{\mathcal{F}}_V = M_V \cdot \vec{g}_V
\end{align}
with the GFF vector $\vec{g}_V=\left(A_{20}(\mathsf{t}),B_{20}(\mathsf{t}),C_{20}(\mathsf{t})\right)^{\intercal}$. The coefficient matrix $M_V$ consists of the
prefactors calculated from \Eq{eq:ffunc} and
$\vec{\mathcal{F}}_V$ is extracted
from a fit of  Eqs.~\eqref{eq:c3pt_a} and \eqref{eq:c3pt_ampl} to lattice data for $C^{2\mathrm{pt}}$ and $C^{3\mathrm{pt}}$.
The number of columns of $M_V$ is equal to the number of unknown GFFs (in this case 3),
but the number of rows depends on the available combinations.
In almost all the cases this yields an overdetermined system of equations, meaning
that the number of elements in $\vec{\mathcal{F}}_V$,
denoted with $\dim{\vec{\mathcal{F}}_V}$,
is larger than the number of GFFs.
Note that the individual rows of $M_V$ are either
real or imaginary.\footnote{If a row vanishes, then it does not
  restrict the GFF and we remove it from the system of equations.}

For a given ensemble this system of equations has to be solved separately for each
virtuality to yield the GFFs as functions of $\mathsf{t}$.
In the general case we write \Eq{eq:eqs_} as
\begin{align}
\label{eq:eqs}
\vec{\mathcal{F}}_{\Gamma}^{\,q} = M_{\Gamma} \cdot \vec{g}_{\Gamma}^{\,q} \,,
\end{align}
where $\Gamma$ can take the values $V$, $A$, $T$
and $\vec{g}_{\Gamma}^{\,q}$ is the vector of the respective GFFs [cf.~Eqs.~\ref{eq:Decompositions}].
Due to equivalent combinations of momenta and polarizations
most rows in the matrix $M_\Gamma$ are equal or differ by a sign only.
We average the corresponding correlation functions, which improves the
signal-to-noise ratio considerably and reduces the number of equations.

\begin{figure}[t]
\centering
\includegraphics[width=0.49\textwidth]{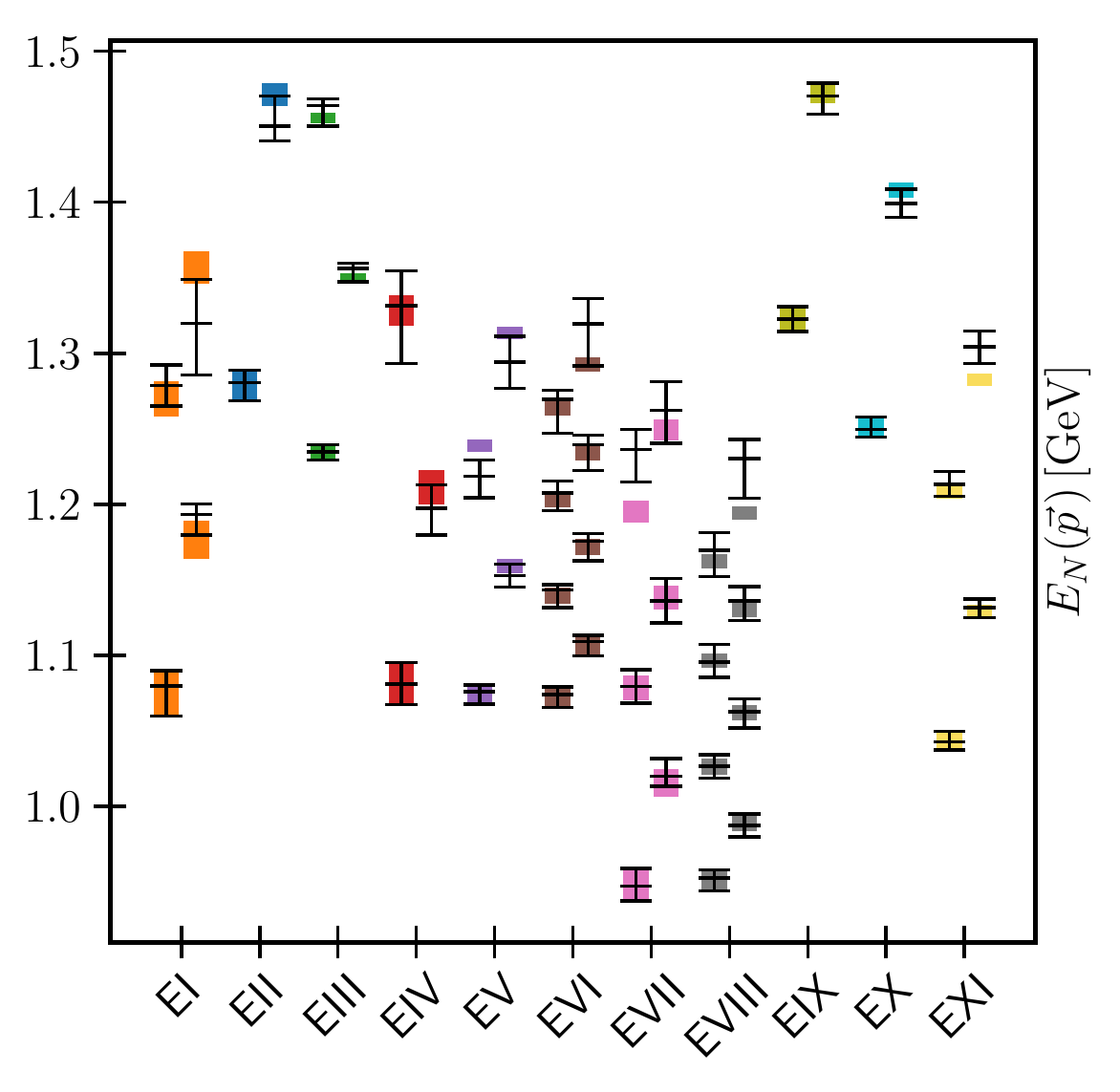}
\caption{
	Overview of the nucleon energies for our ensembles.
	We compare the energies $E_N(\vec{p}\,)$ and the errors extracted from a two-exponential fit
	shown as black error bars
	with the energies $E_N^c$ expected from the continuum dispersion
	relation, which are depicted as colored boxes.
	}
\label{fig:oee}
\end{figure}

\section{Numerical methods}
\label{sec:numerical methods}
\subsection{Gauge ensembles}
\label{sec:gaugens}
Our analysis is based on the large set of gauge configurations produced by the
QCDSF and the RQCD (Regensburg QCD) Collaborations
using the standard Wilson gauge action with two mass-degenerate nonperturbatively
improved clover fermions; see \Tab{tab:latsetup}.
We have three different lattice spacings  0.081\,fm, 0.071\,fm and 0.060\,fm.
Despite the $\mathcal{O}(a)$ improved action,
we expect discretization effects linear in the
lattice spacing for our matrix elements
since the currents are not improved.
The pion masses range from  about 490\,MeV down to 150\,MeV.
In terms of $Lm_\pi$ we cover values from about 3.4 up to 6.7.

\subsection{Fitting two-point correlation functions}
\label{sec:fp2}
\begin{figure}[tb]
\centering
\includegraphics[width=0.48\textwidth]{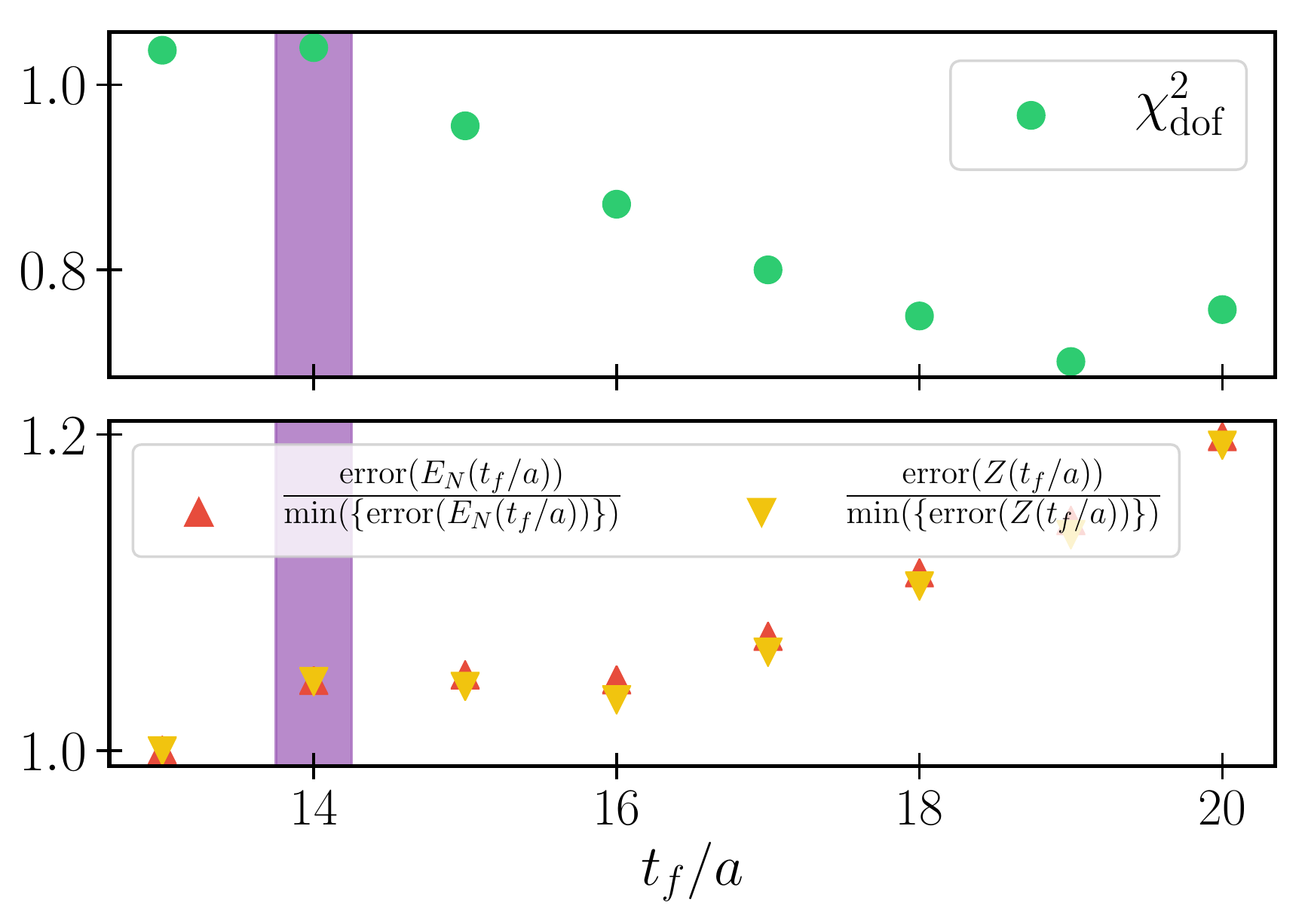}
\caption{The top panel shows the correlated $\chi^2_{\mathrm{dof}}$ as a
function of the final time slice $t_f/a$ for ensemble IV with $E_N(\vec{p}\,)=1.33\,\mathrm{GeV}$;
the bottom panel shows the uncorrelated normalized statistical error of the fit parameters $E_N$ and $Z$.
For the case shown we select the $t_f/a = 14$ result.
}
\label{fig:2ptchoice}
\end{figure}

We parametrize our two-point correlation functions with a two-exponential fit
ansatz
\begin{subequations}
\label{eq:2pt_cor_fit_2e}
\begin{align}
\label{eq:2pt_cor_fit_2e_a}
C^{2\mathrm{pt}}(t , \vec{p}\,)    &=
A(\vec{p}\,) \,   \mathrm{e}^{-E_N(\vec{p}\,) \, t} +
X(\vec{p}\,) \,   \mathrm{e}^{-Y(\vec{p}\,) \, t}
\intertext{with}
\label{eq:2pt_cor_fit_2e_b}
A(\vec{p}\,) &=  Z(\vec{p}\,) \, \frac{E_N(\vec{p}\,) + m_N}{E_N(\vec{p}\,)} ,
\end{align}
\end{subequations}
in order to create bootstrap ensembles for the fit parameters
$A(\vec{p}\,)$, $E_N(\vec{p}\,)$, $X(\vec{p}\,)$ and  $Y(\vec{p}\,)$.
To improve the signal, we average over all
momentum combinations which lead to the same $\vec{p}^{\,2}$.
Subsequently, we use \Eq{eq:2pt_cor_fit_2e_b} to fix the overlap factors
$Z(\vec{p}^{\,\prime})$ and $Z(\vec{p}^{}\,)$
which are needed to factor out $\vec{\mathcal{F}}_\Gamma$
from the three-point correlation functions (cf.~\Eq{eq:c3pt_ampl}).
The fit parameters  $X(\vec{p}\,)$ and  $Y(\vec{p}\,)$ are introduced
in order to parametrize the contributions from excited states.
The parameter $E_N(\vec{p}\,)$ represents the nucleon energy (we do not assume a functional form for the energy).
However, our analysis assumes continuum symmetries.
Therefore  we restrict our lattice calculations to momenta whose fitted values for $E_N(\vec{p}\,)$ are
consistent with the continuum dispersion relation (cf.~Fig.~\ref{fig:oee})
\begin{align}
\label{eq:econt}
  E^c_N(\vec{p}\,) = \sqrt{m_N^2 + \vec{p}^{\, 2}} \, .
\end{align}
The statistical errors are estimated by virtue of 500 bootstrap ensembles.
We carefully study the fit-range dependence of the fit parameters.
Therefore we consider the start time slices $t_s/a \in \{2,3\}$ and vary the final time slice $t_f/a$.
We find that the impact of $t_s/a$ on the values for the GFFs is rather mild
and therefore  we fix $t_s/a = 2$ in the following.
In Fig.~\ref{fig:2ptchoice} we demonstrate how we choose the final time slice $t_f/a$.
We also try single exponential fits and find that they give similar results if
one adjusts the fit ranges appropriately.
However, the resulting errors on $A(\vec{p}\,)$ are larger.
Hence we use the two-exponential fit ansatz for our final analysis.

\subsection{Three-point correlation functions}
\label{sec:fp3}
For the lattice calculations of three-point functions we use the sequential
source method where we set the outgoing nucleon momentum
$\vec{p}^{\,\prime} = \vec{0}$ for all our ensembles.
We parametrize the data using Eqs.~\eqref{eq:c3pt_a}
and \eqref{eq:c3pt_ampl} with $E_N(\vec{p}^{\, \prime})$ = $m_N$.
The initial energy $E_N(\vec{p}^{}\,)$ is determined from the
continuum dispersion relation (\ref{eq:econt}).
The momentum restriction, which we discussed in the previous section,
translates to a range $0 \le  Q^2 < 0.6\, \mathrm{GeV}^2$
for the three-point functions.
With $Z(\vec{p}^{\,\prime})$ and $Z(\vec{p}^{}\,)$ having been determined
from the two-point correlation functions,
the only free parameter left is $\mathcal{F}^{q}_\Gamma$.
To achieve ground state dominance,
one has to make sure that
$aN_T \gg  t^{\prime} \gg \tau \gg 0$ [cf.\ \Eq{eq:c3pt_a}].
We consider $\tau\in [\tau_s,\tau_e]$ where
$\tau_s$ is well above zero and $\tau_e$ well below $t^\prime$.
The sink times vary with the ensemble
(see the last column of \Tab{tab:latsetup}).
In Sec.\,\ref{sec:error_estimation} we examine possible excited state contaminations.

\subsection{Determination of the GFFs}
\label{sec:chi_gff}
As explained above, for every
current $\Gamma=V$, $A$ or $T$, quark flavor $q$ and virtuality
$-\mathsf{t}$, we need to solve the linear system \Eq{eq:eqs}, i.e.,
$\vec{\mathcal{F}} = M \cdot \vec{g}$, to extract the
relevant form factors $\vec{g}$ from the vector of inequivalent
matrix elements $\vec{\mathcal{F}}$ that correspond to nonvanishing
rows of $M$. Here
we drop all indices like the quark flavor $q$ and $\Gamma$ for convenience.
In what follows $m$ denotes the number of independent
form factors while $n\geq m$ is the length of $\vec{\mathcal{F}}$.
Consequently, $M$ is a $n\times m$ matrix of maximal rank,
i.e., $\rank(M)=m$.

The determination of the form factors is carried out in two ways.
The first method consists of two steps: First we extract the
ground state nucleon matrix elements $\mathcal{F}_j$ from the
lattice three-point function data $c^\tau_j$, restricted to the
range of insertion times $\tau\in[\tau_s,\tau_e]$,
through the numerical minimization of the $\chi^2$-function
\begin{equation}
 \label{eq:chi}
 \chi^2\big(\vec{\mathcal{F}}\,\big) =
 \sum\limits_{j = 1}^{n}
 \sum\limits_{\tau, \tau^\prime =\tau_s }^{\tau_e}
 \delta c_{j}^{\tau}
	 \left[
		 \mathrm{cov}^{-1}_j
	 \right]_{\tau\tau^\prime}
 \delta c_{j}^{\tau^\prime} \,,
\end{equation}
where $\delta c_j^\tau$ is the difference
\begin{equation}
 \label{eq:delta}
 \delta c_{j}^{\tau} = c_{j}^{\tau} - \mathcal{F}_j\sqrt{ Z(\vec{p}^{\,\prime}) Z(\vec{p}^{}\,)}
    \;\mathrm{e}^{-m_N(t^{\prime} - \tau)} \, \mathrm{e}^{-E_N(\vec{p}\,)\tau}
\end{equation}
between the lattice data and the three-point function
parametrization \Eq{eq:c3pt_ampl}.
The inverse covariance matrix $\mathrm{cov}_j^{-1}$ depends on the
insertion times $\tau$ and $\tau^\prime$. One can easily generalize
the fit to the situation of multiple source-sink distances $t'$ if this
is required or include excited state contributions.
The index $j\in\{1, \ldots, n\}$ runs over all possible polarizations $\rho$
and initial momenta $\vec{p}$ (keeping the virtuality $Q^2$ fixed),
which give nonvanishing contributions.

Once the fit parameters $\mathcal{F}_j$ are determined,
one can minimize
\begin{equation}
\label{eq:bad}
 \epsilon^2   =
 \left( M \vec{g} - \vec{\mathcal{F}} \right)^2
\end{equation}
to determine the form factors $\vec{g}$.
The total number of parameters for this method is $m+n$ and,
in particular for large virtualities, this number can be quite large (up to 50).
This is not the only problem but
it can happen that the resulting $\epsilon$ value is quite large and
it is not clear how one should deal with such a situation.

\begin{figure}[t]
    \centering
   \includegraphics[width=0.47\textwidth]{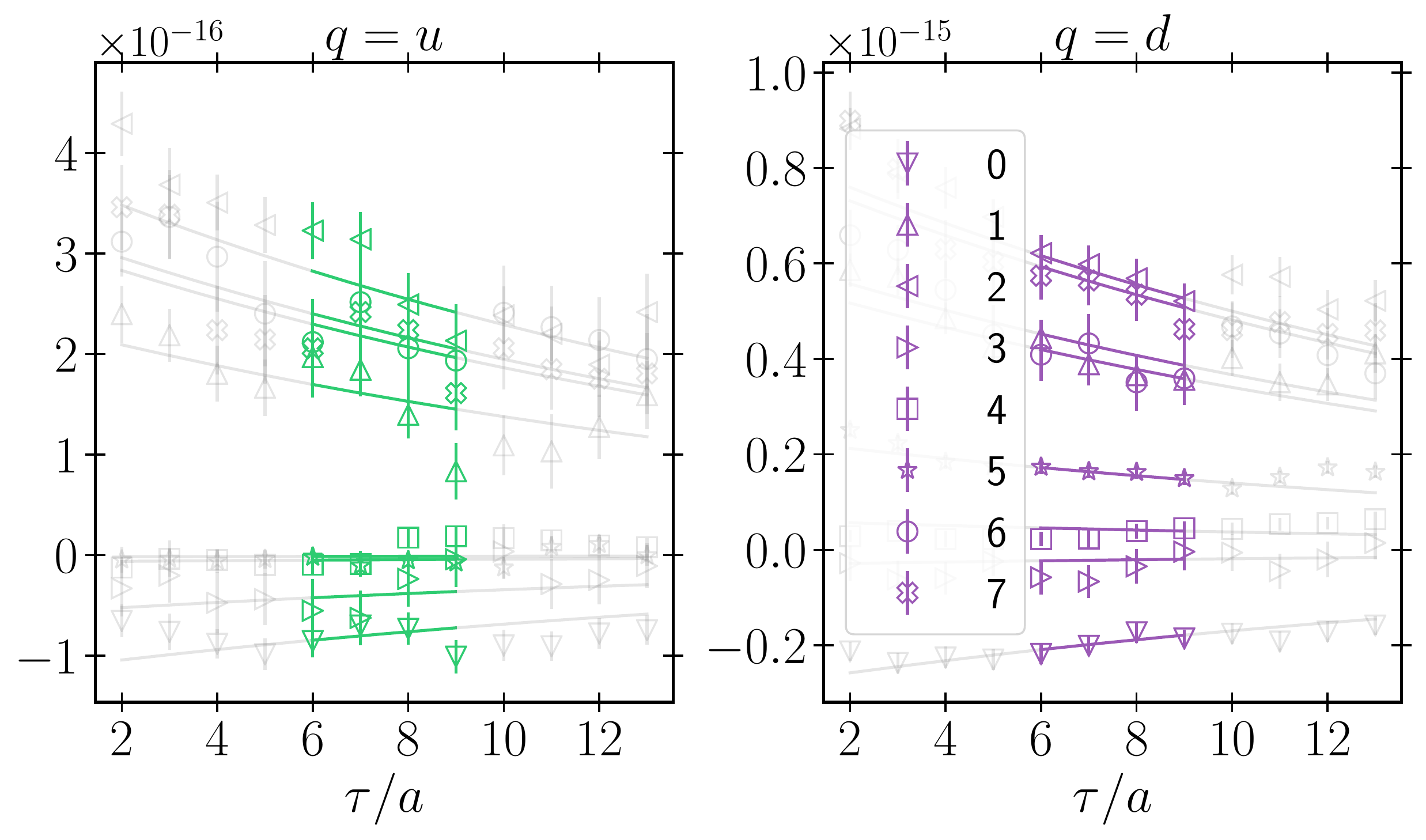}
   \caption{
     Fit results using the single step minimization method.
     We show ensemble VIII at the virtuality $Q^2 = 0.277\,\mathrm{GeV}^2$
     in the vector channel. This corresponds to a
     spatial momentum transfer of $2\cdot 2\pi/L$, where we
     have averaged over all equivalent lattice directions.
     Three fit parameters
     $\vec{\alpha} = (\alpha_1, \, \alpha_2, \, \alpha_3)^{\intercal}$
     fully describe eight three-point functions.
     Colored points lie in the fit range $[\tau_s,\tau_e]$ [cf.\ \Eq{eq:chi}].
     On the left we show data for the $u$ quark and on the right
     for the $d$ quark~(omitting disconnected contributions). The numbers in the legend
       refer to the channels listed in
       \Tab{tab:operatorcontributions_to_fit}.
   \label{fig:imag_space}}
\end{figure}

Ideally, $\epsilon$ should be zero but this is only possible if
$\vec{\mathcal{F}}$ is in the image of $M$ [cf.~Eq.~\eqref{eq:bad}].
Motivated by this observation, we carry out our fits employing a single step
method, which combines the two subsequent steps into a single
minimization problem, restricting the number of fit parameters to the
relevant degrees of freedom.
We start from the singular value decomposition,
\begin{equation}
	 \label{eq:SVD}
	 M = U  \cdot \Sigma \cdot V^{\intercal}
\end{equation}
with orthogonal matrices
$U\in\mathbb{R}^{n \times n}$, $V\in \mathbb{R}^{m \times m}$
and the matrix $\Sigma\in \mathbb{R}^{n \times m}$, which
has nonvanishing entries only on the diagonal. The pseudoinverse $\Sigma^+$ is
a $m\times n$ matrix that can easily be
obtained, computing the inverses of the diagonal elements of $\Sigma$.
Each vector $\vec{\mathcal{F}}$
within the image of $M$ can be uniquely expressed
as a linear combination
\begin{equation}
	\label{eq:imag}
        \vec{\mathcal{F}}(\vec{\alpha}\,) = \sum\limits_{i = 1}^{m} \alpha_i \, \vec{u}^{\,i}
\end{equation}
of the first $m$ column vectors of $U$. Note that $m = \rank({M})$.
Substituting $\vec{\mathcal{F}} \mapsto \vec{\mathcal{F}}(\vec{\alpha}\,)$
in Eq.~\eqref{eq:delta} [and thereby Eq.~\eqref{eq:chi}],
we obtain a modified $\chi^2$-function that depends on the
parameters $\alpha_i$, where $i\in\{1, \ldots, m\}$.
Finally, we convert the extracted vector $\vec{\alpha}$ to the
desired GFF vector,
\begin{align}
\label{eq:SVDI}
\vec{g} =
\left[V\Sigma^+ U^{\intercal}\right]\sum\limits_{i = 1}^{m} \alpha_i \, \vec{u}^{\,i} =
 \left[V\Sigma^+\right]\vec{\alpha}\,,
\end{align}
where in the last step $\Sigma^+$ is truncated to a $m\times m$ square matrix.
In \Fig{fig:imag_space} we show for one example on the nearly physical
quark mass ensemble VIII that this method works very well. In this
case eight different lattice channels, listed in
  \Tab{tab:operatorcontributions_to_fit}, are well described in terms
of three fit parameters.

\begin{table}[t]
    \caption{Individual operator contributions to the fits shown
      in \Fig{fig:imag_space}.
      The numbers in the legend of \Fig{fig:imag_space} correspond to the
      channels below.
      We parametrize the spatial lattice momentum $\vec{q}= \hat{k}2\pi/L$ 
      in terms of $\hat{e}_1$,$\hat{e}_2$, and $\hat{e}_3$ which are unit vectors in the three spatial directions. \label{tab:operatorcontributions_to_fit}
} 
  \begin{center} \begin{ruledtabular}
        \begin{tabular}{l@{\quad}c@{\quad}c@{\qquad}c@{\qquad} c@{\qquad}  c@{\qquad}   }
            Channel       &  $\mathbb{P}^\rho$   & $\mathcal{O}$                    &$\hat{k}$          & Channel & \#contrib.              \\
            \hline
            0         &   $\mathbb{P}^4$     & $\mathcal{O}_{1\,4}^{v_{2,a}}$    & $\pm2\hat{e}_1$ & imaginary &  2 \\
                      &                      & $\mathcal{O}_{2\,4}^{v_{2,a}}$    & $\pm2\hat{e}_2$ &  &  2   \\
                      &                      & $\mathcal{O}_{3\,4}^{v_{2,a}}$    & $\pm2\hat{e}_3$ &  &  2   \\
            \hline
            1         &                      & $\mathcal{O}_{1}^{v_{2,b}}$       & $\pm2\hat{e}_1$ & real &  2 \\
                      & &                                                        & $\pm2\hat{e}_2$ &     &  2  \\
            \hline
            2         &                      & $\mathcal{O}_{2}^{v_{2,b}}$       & $\pm2\hat{e}_1$ & real&  2  \\
                      & &                                                        & $\pm2\hat{e}_2$ &     &  2  \\
            \hline
            3         &                      & $\mathcal{O}_{3}^{v_{2,b}}$       & $\pm2\hat{e}_1$ & real&  2  \\
                      & &                                                        & $\pm2\hat{e}_2$ &      &  2 \\
            \hline
            4        &   $\mathbb{P}^1$     & $\mathcal{O}_{2\,3}^{v_{2,a}}$     & $\pm2\hat{e}_2$, $\pm2\hat{e}_3$ &  imaginary   &  4  \\
                     &   $\mathbb{P}^2$     & $\mathcal{O}_{1\,3}^{v_{2,a}}$     & $\pm2\hat{e}_1$, $\pm2\hat{e}_3$ &       &  4   \\
                     &   $\mathbb{P}^3$     & $\mathcal{O}_{1\,2}^{v_{2,a}}$     & $\pm2\hat{e}_1$, $\pm2\hat{e}_2$ &       &  4   \\
            \hline
            5        &   $\mathbb{P}^1$     & $\mathcal{O}_{3\,4}^{v_{2,a}}$     & $\pm2\hat{e}_2$&  real   &  2  \\
                     &                      & $\mathcal{O}_{2\,4}^{v_{2,a}}$     & $\pm2\hat{e}_3$&         &  2  \\
                     &   $\mathbb{P}^2$     & $\mathcal{O}_{3\,4}^{v_{2,a}}$     & $\pm2\hat{e}_1$&         &  2  \\
                     &                      & $\mathcal{O}_{1\,4}^{v_{2,a}}$     & $\pm2\hat{e}_3$&         &  2  \\
                     &   $\mathbb{P}^3$     & $\mathcal{O}_{2\,4}^{v_{2,a}}$     & $\pm2\hat{e}_1$&         &  2  \\
                     &                      & $\mathcal{O}_{1\,4}^{v_{2,a}}$     & $\pm2\hat{e}_2$&         &  2  \\
            \hline
            6        &   $\mathbb{P}^4$     & $\mathcal{O}_{1}^{v_{2,b}}$        & $\pm2\hat{e}_3$&  real   &  2  \\
            \hline
            7        &   $\mathbb{P}^4$     & $\mathcal{O}_{2}^{v_{2,b}}$        & $\pm2\hat{e}_3$&  real   &  2  \\
      \end{tabular}
      \end{ruledtabular}
  \end{center}
\end{table}

\begin{figure}[b]
    \centering
   \includegraphics[width=0.47\textwidth]{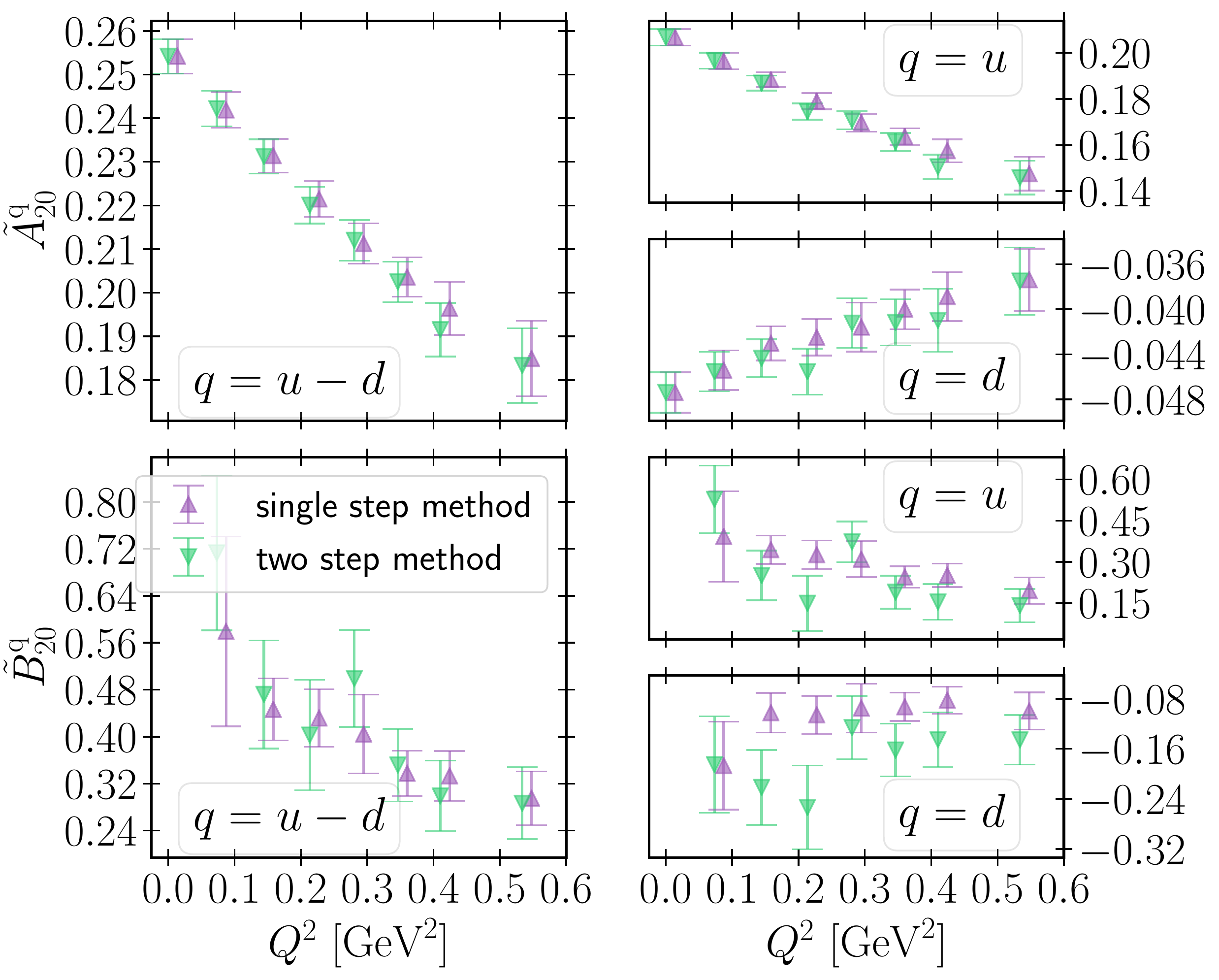}
   \caption{
   Comparison of single step and two step fit methods for the axial GFFs for ensemble VI. The right panels show $\widetilde{A}_{20}$ and $\widetilde{B}_{20}$ separately for the $u$ and $d$ quark (without disconnected contributions), the left panels for the isovector case.
   \label{fig:comp_im_da}}
\end{figure}

\begin{figure}[t]
   \includegraphics[width=0.47\textwidth]{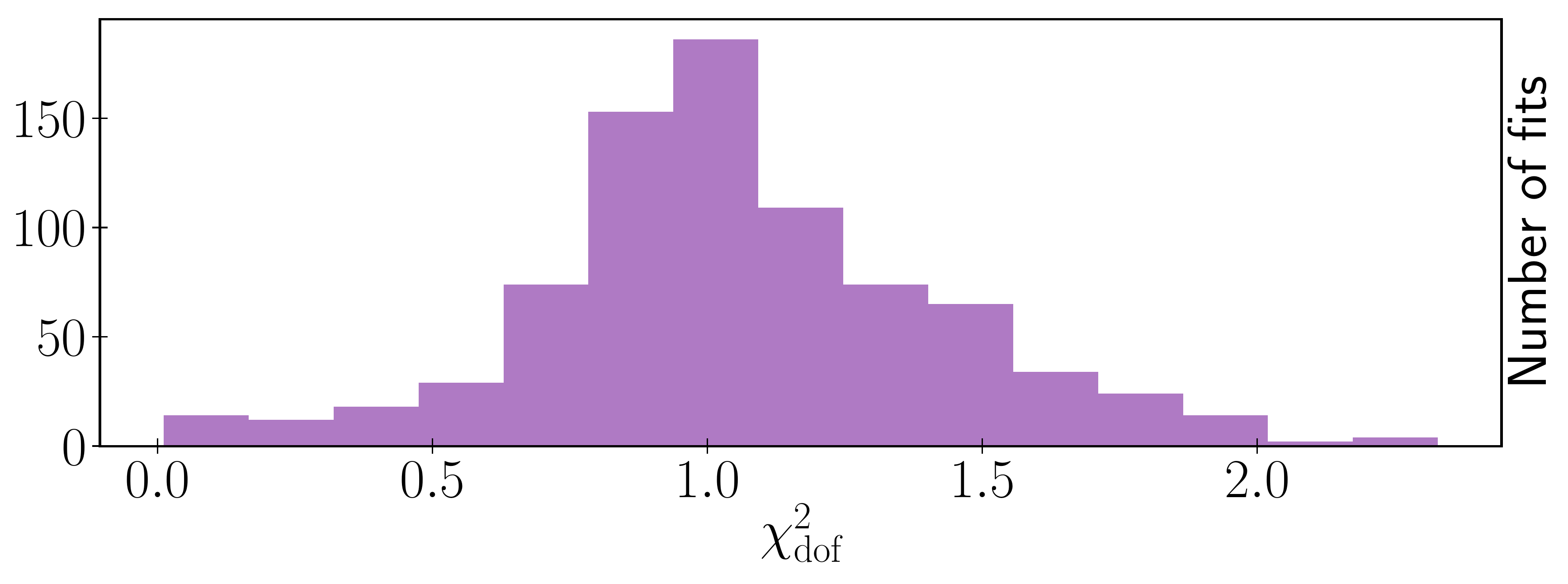}
   \caption{$\chi^2$ distribution of all GFF fits performed for this analysis.}
\label{fig:chi2}
\end{figure}

A comparison of the two fit methods shows that the results
are consistent within errors for all GFFs and for all ensembles.
The single step method, however, results in somewhat
smaller statistical errors and a
smoother $Q^2$ dependence, especially for the induced GFFs.
In \Fig{fig:comp_im_da} we directly compare the two methods.
For the final results we only use the single step method.
In \Fig{fig:chi2} we show all $\chi^2_{\mathrm{dof}}$ values of all fits used
in this paper to extract all considered GFFs:
The correlated single step
fits provide a very satisfactory description of the data.

\subsection{Excited states}
\label{sec:error_estimation}
For some of our ensembles we have three-point function data for
different source-sink separations.
This allows us to analyze the influence of excited states on the GFFs.
Our analysis is based on ensemble IV with five source-sink separations in the
range $t^\prime/a \in [7, 17]$ and on ensemble VIII
with three source-sink separations in the range $t^\prime/a \in [9, 15]$.
In physical units $t^\prime = 15a$ corresponds to about $1\,\mathrm{fm}$.
Ensemble VIII  has data for eight values of $Q^2$ and this
ensemble corresponds to an almost physical pion mass.
We show results only for this ensemble, but our findings are consistent for both ensembles.

\begin{figure}[t]
    \centering
    \includegraphics[width=0.45\textwidth]{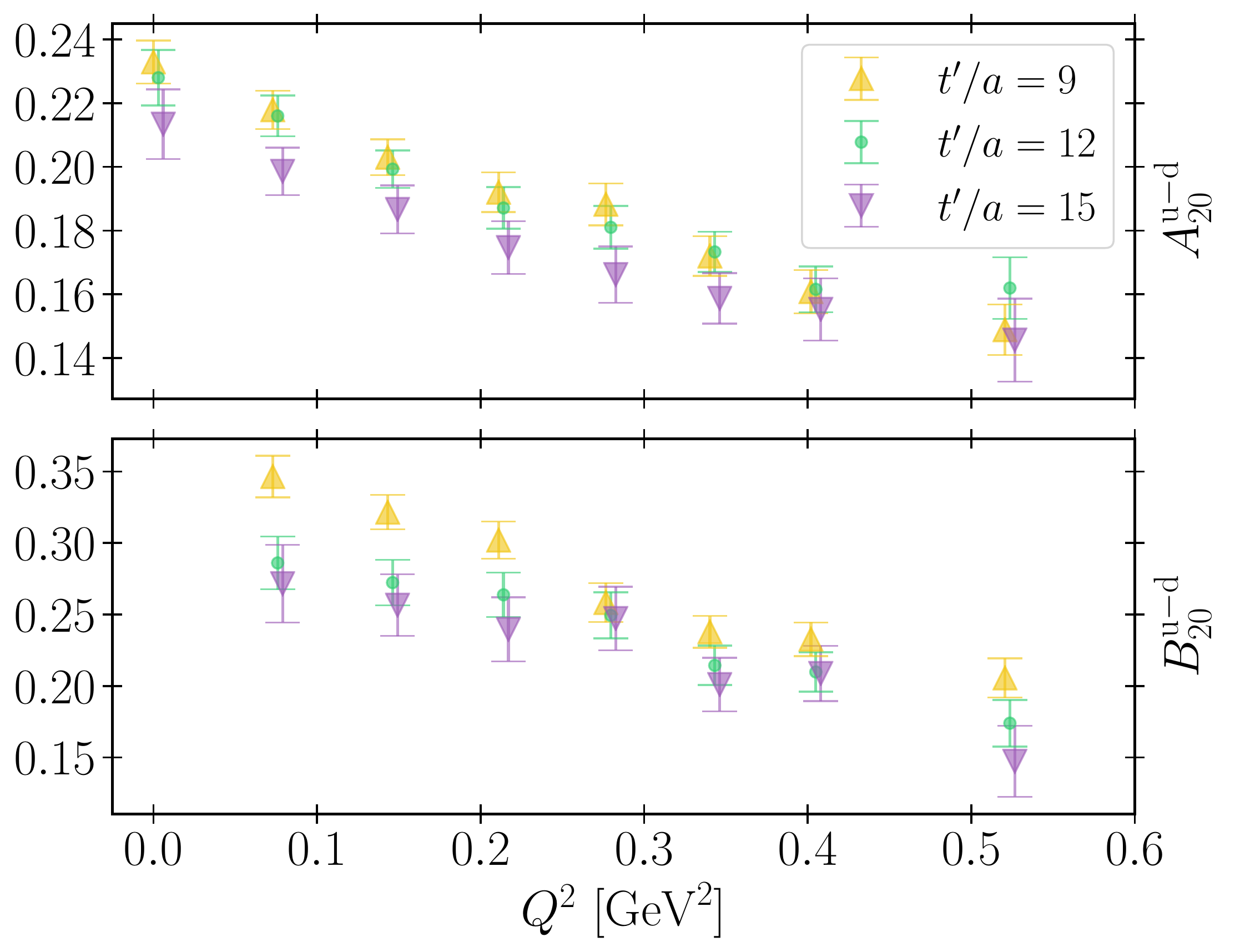}
   \caption{The vector GFFs vs $Q^2$ for different sink
   times $t^{\prime}$ for ensemble VIII. \label{fig:vGFFx}
   }
\end{figure}

For the tensor and axial GFFs we find that within statistical
errors the $Q^2$ dependence is not affected by a
variation of $t^\prime$.
Only in the vector case, especially for $A_{20}^{u-d}$,
excited state contaminations are visible (see \Fig{fig:vGFFx}).
We have tried to parametrize these excited-state contributions to the
three-point function with various multiexponential fit ans{\"a}tze.
This, however, introduces additional fit parameters,
in particular the mass and the energy of the first excited state.
The first excitation in the three-point function can be a multihadron state
and hence its energy will in general not be well approximated
by the single particle continuum dispersion relation.
To parametrize excited state contributions clearly several source-sink
separations are required. However, within present statistical
errors little movement is visible for $t'\gtrsim 0.9\,$fm, even
in the $A_{20}$ channel where we achieve the highest accuracy;
see \Fig{fig:vGFFx} for an example.
We therefore have restricted our GFF fits to ranges of $\tau$
where the data
are well described by a single exponential (cf.~Fig.~\ref{fig:imag_space}).
In all the cases $t'$ is larger than 1\,fm.

\begin{figure*}
    \centering
\includegraphics[width=0.94\textwidth]{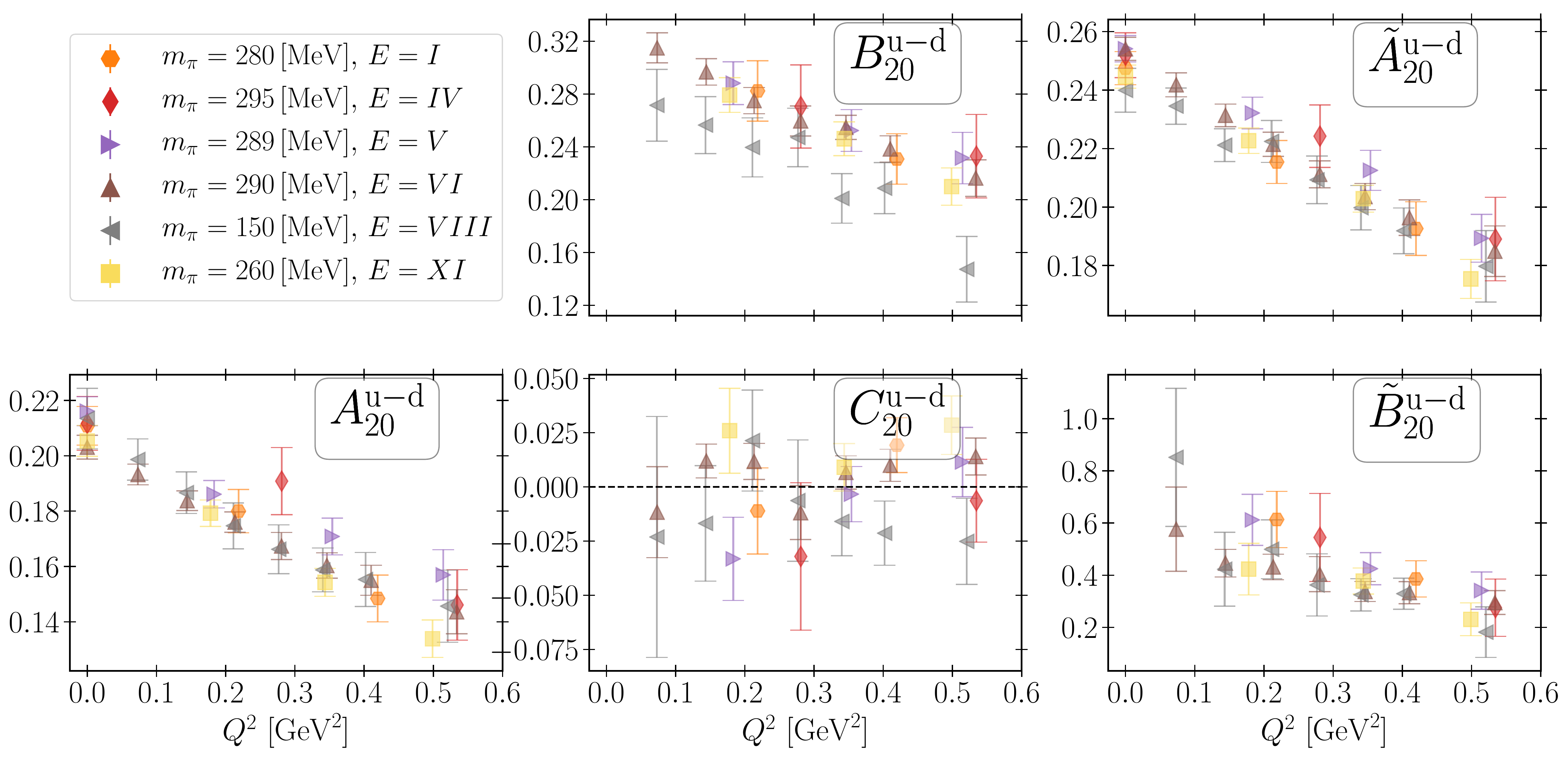}
   \caption{The vector and axial GFFs vs $Q^2$. Left: $A_{20}^{u-d}$,
   $B_{20}^{u-d}$ and $C_{20}^{u-d}$; right: $\widetilde{A}_{20}^{u-d}$ and $\widetilde{B}_{20}^{u-d}$. All results are for the isovector case and in the $\MS$ scheme ($\mu=2\,\text{GeV}$).
  \label{fig:vaGFF}}
\bigskip
   \includegraphics[width=0.94\textwidth]{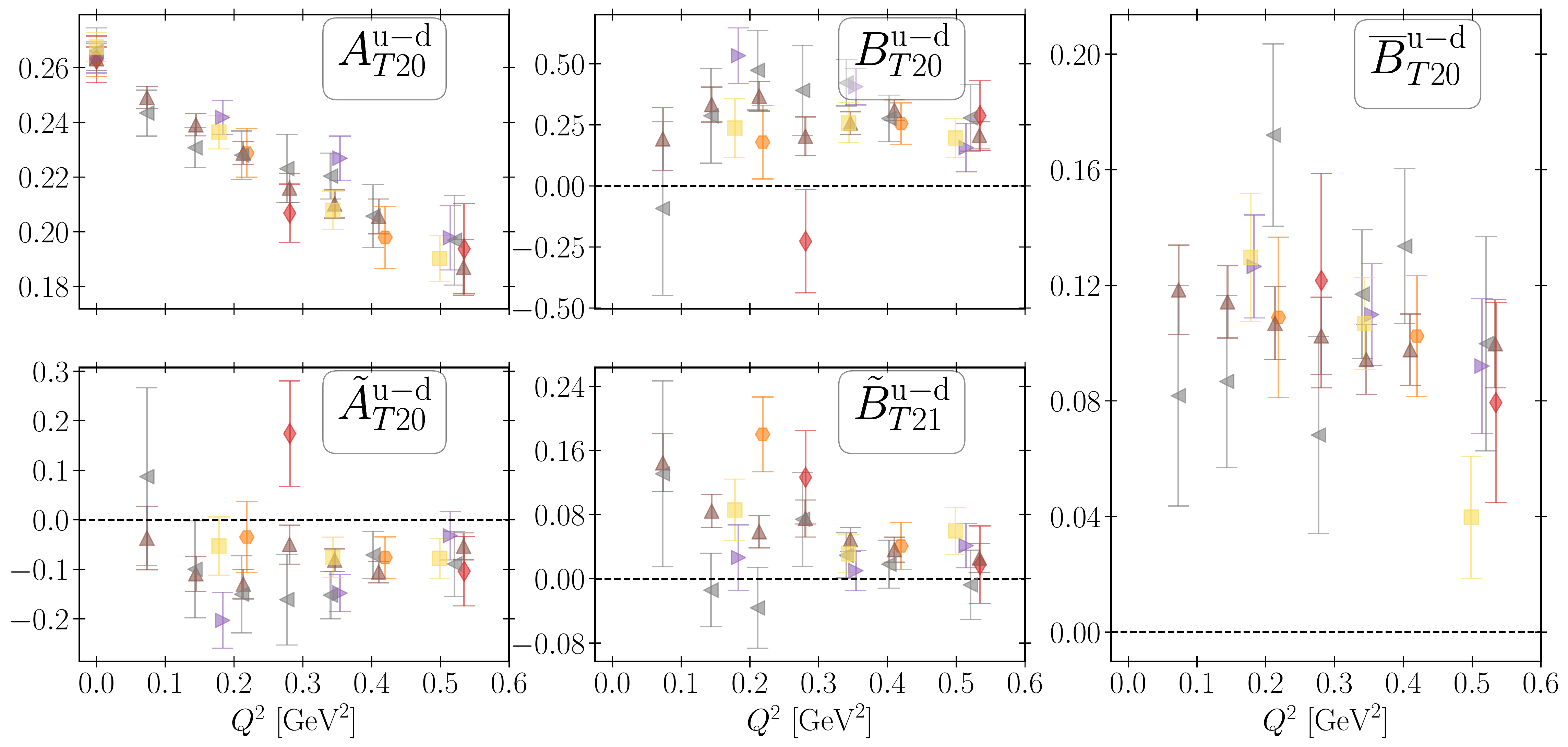}
   \caption{
     The tensor GFFs
     $A^{u-d}_{T20}$,
   $B^{u-d}_{T20}$,
   $\widetilde{A}^{u-d}_{T20}$,
   $\widetilde{B}^{u-d}_{T21}$
   and the linear combination
   $\overline{B}^{u-d}_{T20}$ in
   the $\MS$ scheme ($\mu=2\,\text{GeV}$).
   \label{fig:tGFF}}
\end{figure*}

\section{Nucleon GFFs}
\label{sec:results_GFF}
Below we show results for the nucleon GFFs on a subset of the
ensembles listed
in \Tab{tab:latsetup}. We restrict ourselves to $m_{\pi}<300\,$MeV and
$m_\pi L>3.4$ and analyze the quark mass, volume and lattice spacing
dependence.
All results refer to the $\MS$ scheme at $\mu=2\,\text{GeV}$.

\subsection{Vector and axial GFFs}
Results for the vector GFFs, $A_{20}^{u-d}$, $B_{20}^{u-d}$ and
$C_{20}^{u-d}$, are shown in \Fig{fig:vaGFF} (left) as a function of
$Q^2=-\mathsf{t}$.  We see that the discretization effects are
negligible within errors (comparing ensembles I and XI, which give
about the same pion mass and a similar value for $Lm_\pi$). Also the
volume dependence (cf.\ V and VI) is small, although there is a slight
trend towards larger values for $B_{20}^{u-d}$ if $Lm_\pi$ increases
from about 4.2 to 6.7. For $A_{20}^{u-d}$ and $C_{20}^{u-d}$ we do not
see any volume dependence within present errors.  Similar statements
hold for the quark mass dependence: For $A_{20}^{u-d}$ and
$C_{20}^{u-d}$ it is negligible within errors, but for $B_{20}^{u-d}$
we see a trend towards lower values if the pion mass decreases down to
150 MeV (cf.\ VIII and VI). However, the latter could also be a volume
artifact, since there is also a clear correlation between $Lm_\pi$ and
$B_{20}^{u-d}$ (cf.\ ensembles VIII, V and VI where $Lm_\pi\simeq3.5$,
4.2 and 6.7, respectively).  $A_{20}^{u-d}$ and $B_{20}^{u-d}$ have a
roughly linear $Q^2$ dependence for small $Q^2$, and $C_{20}^{u-d}$ is zero
within errors. This agrees with the leading $\mathsf{t}$-dependence
expected from covariant baryon chiral perturbation theory (BChPT, see
below). We remark that also the individual (quark line
  connected) $u$ and $d$ quark contributions to $C_{20}^{u-d}$ are zero
  within error. So the smallness of this generalized form factor is
  not due to an approximate cancellation.
For large $Q^2$ we expect that $A_{20}^{u-d}$ exhibits a
dipolelike $Q^2$-dependence, which we saw in our former study
(cf.\ Fig.~2 of Ref.~\cite{Sternbeck:2012rw}).

Results for the axial GFFs are shown in the right panel of
\Fig{fig:vaGFF}. We see that a change of volume, quark mass or lattice
spacing has almost no effect on the data. Within errors these effects
cannot be resolved.  Both form factors grow approximately linearly for
$Q^2\to 0$. For $\widetilde{B}_{20}^{u-d}$ the statistical errors
become larger for $Q^2 \rightarrow 0$ whereas the errors for
$\widetilde{A}_{20}^{u-d}$ are nearly independent of $Q^2$.

\subsection{Tensor GFFs}
Continuing with the tensor GFFs,
we show results for $A^{u-d}_{T20}$,  $B^{u-d}_{T20}$,
$\widetilde{A}^{u-d}_{T20}$ and
$\widetilde{B}^{u-d}_{T21}$ in \Fig{fig:tGFF}.
The dominant form factors are $A^{u-d}_{T20}$ and $B^{u-d}_{T20}$.
For the available virtualities $A^{u-d}_{T20}$ rises linearly for $Q^2\to 0$,
while $B^{u-d}_{T20}$ remains more or less constant, well above zero.
Overall, the statistical errors for $A^{u-d}_{T20}$ are smaller than for $B^{u-d}_{T20}$.
Volume, quark mass or lattice spacing effects cannot be resolved within errors.

\begin{figure}[b]
    \centering
    \includegraphics[width=0.45\textwidth]{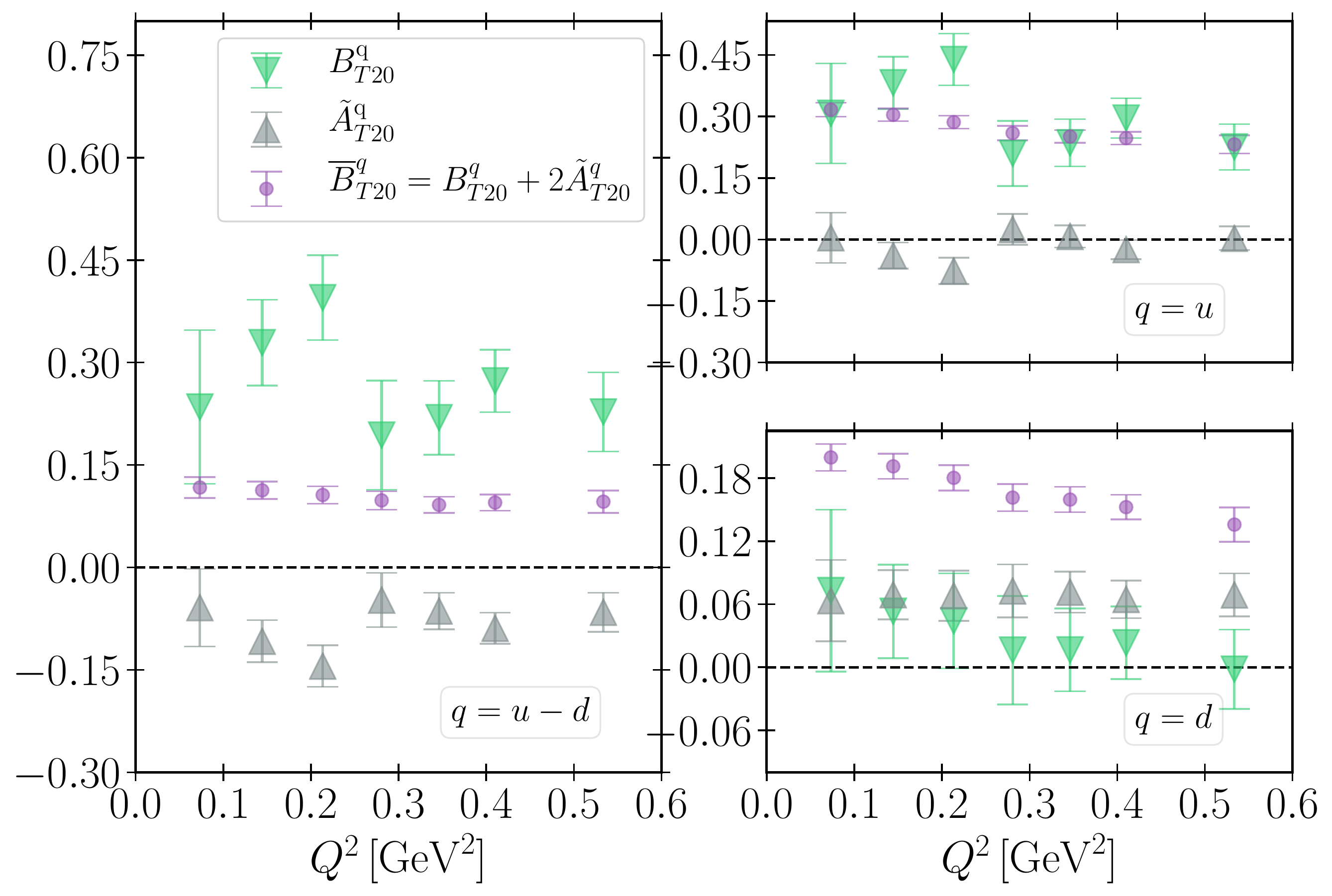}
   \caption{
   Strong anticorrelations between $B^q_{T20}$ and  $\widetilde{A}^q_{T20}$
   for the example of ensemble VI.
   \label{fig:tensor_anti}
   }
\end{figure}

The other two GFFs, $\widetilde{A}^{u-d}_{T20}$ and $\widetilde{B}^{u-d}_{T21}$,
are smaller in comparison and,
besides a few outliers, are best described by a constant.
However, a final conclusion cannot be drawn as
the statistical errors for both GFFs are rather large.
We also study the linear combination
\begin{align}
 \label{eq:Bbar}
 \overline{B}^{q}_{T20} =  B^{q}_{T20} + 2\widetilde{A}^{q}_{T20}\,,
\end{align}
which corresponds to the combination of GPDs $E_T  + 2\tilde{H}_{T}$ that is
related to
the Boer-Mulders function $h_1^\perp$~\cite{Boer:1997nt}.
We find that the statistical error of $\overline{B}^{q}_{T20} $ is
significantly smaller compared to the individual errors of $B^{q}_{T20}$
and $\widetilde{A}^{q}_{T20}$ (see \Fig{fig:tensor_anti}).
We will take advantage of this observation when looking at the transverse
spin of the nucleon in Sec.\,\ref{sec:nucleon_tomography}.
The results for $\overline{B}^{u-d}_{T20}$ are shown with the
tensor GFFs in \Fig{fig:tGFF}
for the same ensembles.
The anticorrelations we find for $\overline{B}^{u-d}_{T20}$ are
present for all ensembles.

\section{Extraction of \texorpdfstring{$\boldsymbol{J^{u-d}}$}{J(u-d)}}
\label{sec:chiral_J}
The GFFs $A_{20}^{u-d}(\mathsf{t})$ and $B^{u-d}_{20}(\mathsf{t})$ are of particular interest since for $\mathsf{t}\to0$ they
are related to the total angular momentum~\cite{Ji:1996ek}
\begin{align}
  \label{eq-ji}
  J^{u-d}  = \frac{1}{2} \left[A_{20}^{u-d}(0) + B_{20}^{u-d} (0) \right].
\end{align}
In order to estimate $J^{u-d}$ at the physical pion mass we analyze our data for $A_{20}^{u-d}(\mathsf{t})$ and
$B^{u-d}_{20}(\mathsf{t})$, employing the BChPT formulas of Ref.~\cite{Wein:2014wma}, which, however, we truncate at order $m_{\pi}^3$,
\begin{align}
\label{eq:chia}
 A^{u-d}_{20}(\mathsf{t}, m_\pi)  = \left[ 1 - \frac{(1 + 3\,g_A^2) \, m_\pi^2 \, \log( \frac{m_\pi^2}{\mu^2}) }{16\, f_\pi^2\, \pi^2} \right] \, L \, \nonumber \\
  +\, m_\pi^2 \, M_2^A \, + m_\pi^3\,M_3^A\,+ \mathsf{t}(\, T_0^A + m_\pi^2 \, T_1^A)
\end{align}
and
\begin{align}
\label{eq:chib}
&B^{u-d}_{20}(\mathsf{t}, m_\pi) =
\frac{ g_A^2 \,  m_\pi^2\, \log(\frac{m_\pi^2}{ \mu^2}) }{16\, f_\pi^2 \,\pi^2}L
+ \, \mathsf{t}(\, T_0^B + m_\pi^2 \, T_1^B) \, \nonumber \\
&+\left[ 1 - \frac{(1 + 2 \,  g_A^2) \,m_\pi^2 \log(\frac{m_\pi^2}{\mu^2})}{16 \, f_\pi^2 \, \pi^2}\right] L^B \, +m_\pi^2 \, M_2^B \,.
\end{align}
The fit parameters $T_1^A$ and $T_1^B$ are added since our data
extend up to virtualities
$-\mathsf{t} \approx (770\,\mathrm{MeV})^2 \gg m_{\pi}^2$,
however, these terms would naturally appear at the next order of BChPT.
We determine the parameters $(L, M_2^A, M_3^A, T_0^A, T_1^A)$
and $(L, L^B, M_2^B, T_0^B, T_1^B)$ by carrying out
combined fits to our data sets for
$A^{u-d}_{20}(\mathsf{t},\,  m_\pi)$ and
$B^{u-d}_{20}(\mathsf{t},\, m_\pi)$.
The remaining parameters in Eqs.~\eqref{eq:chia} and \eqref{eq:chib}
are constrained to \hbox{$g_A = 1.256$}, \hbox{$f_\pi = 92.4\,\mathrm{MeV}$} and \hbox{$\mu = 1.0 \,\mathrm{GeV}$}.

\begin{figure}[tb]
  \centering
  \includegraphics[width=0.45\textwidth]{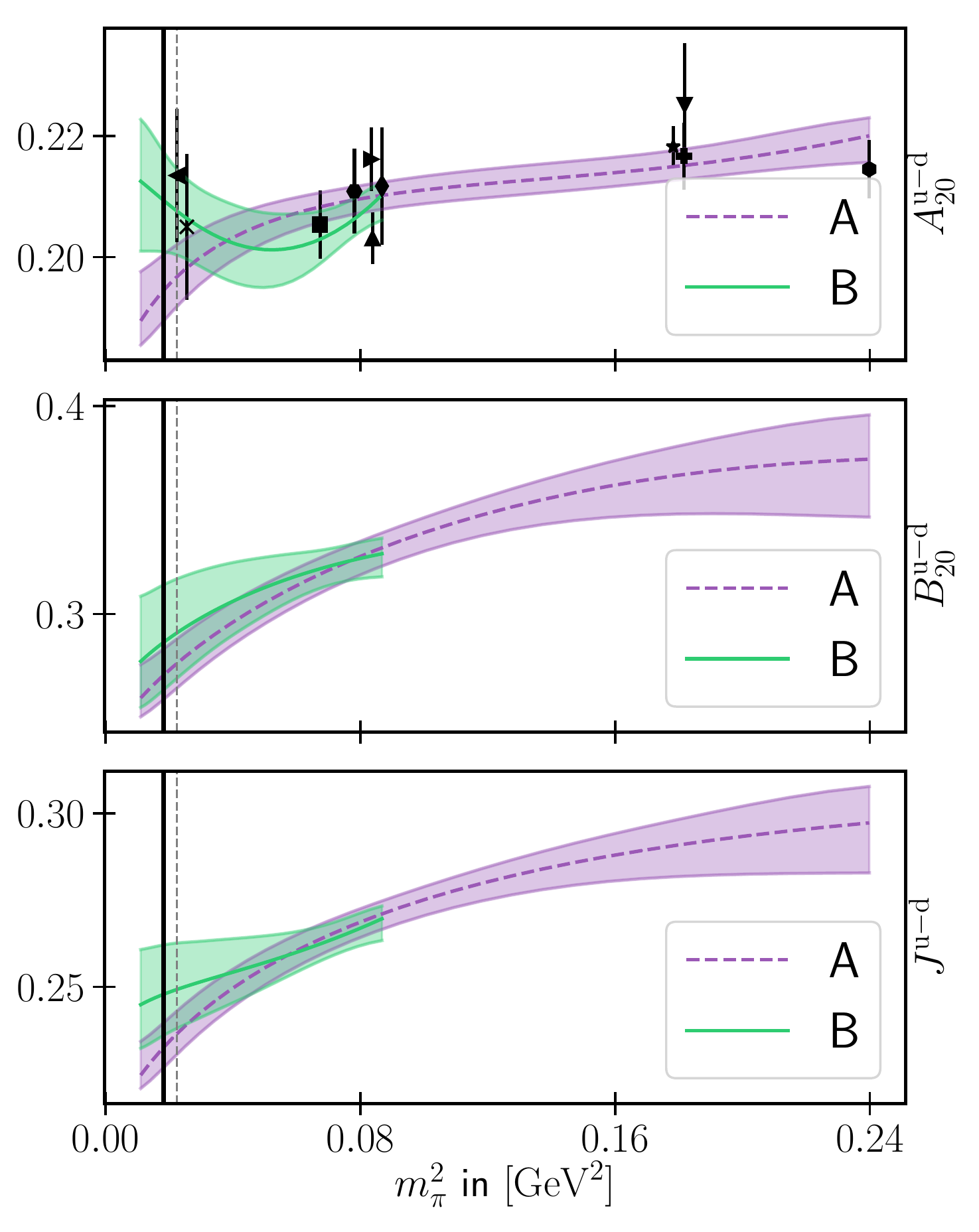}
  \caption{
  From top to bottom $A_{20}^{u-d}(0)$, $B_{20}^{u-d}(0)$ and $J^{u-d}$ as a function of the pion mass squared.
  The vertical solid line marks the physical pion mass; the vertical dashed line
  indicates our smallest pion mass. The A-band is from a fit of all our ensembles and the B-band
  from a fit where ensembles with $m_\pi > 300 \,\mathrm{MeV}$ are removed. For $A_{20}^{u-d}(0)$ we
  have lattice data which are shown in the top panel for comparison.
  \label{fig:j}
  }
\end{figure}

\begin{figure}[tb]
  \centering
  \includegraphics[width=0.45\textwidth]{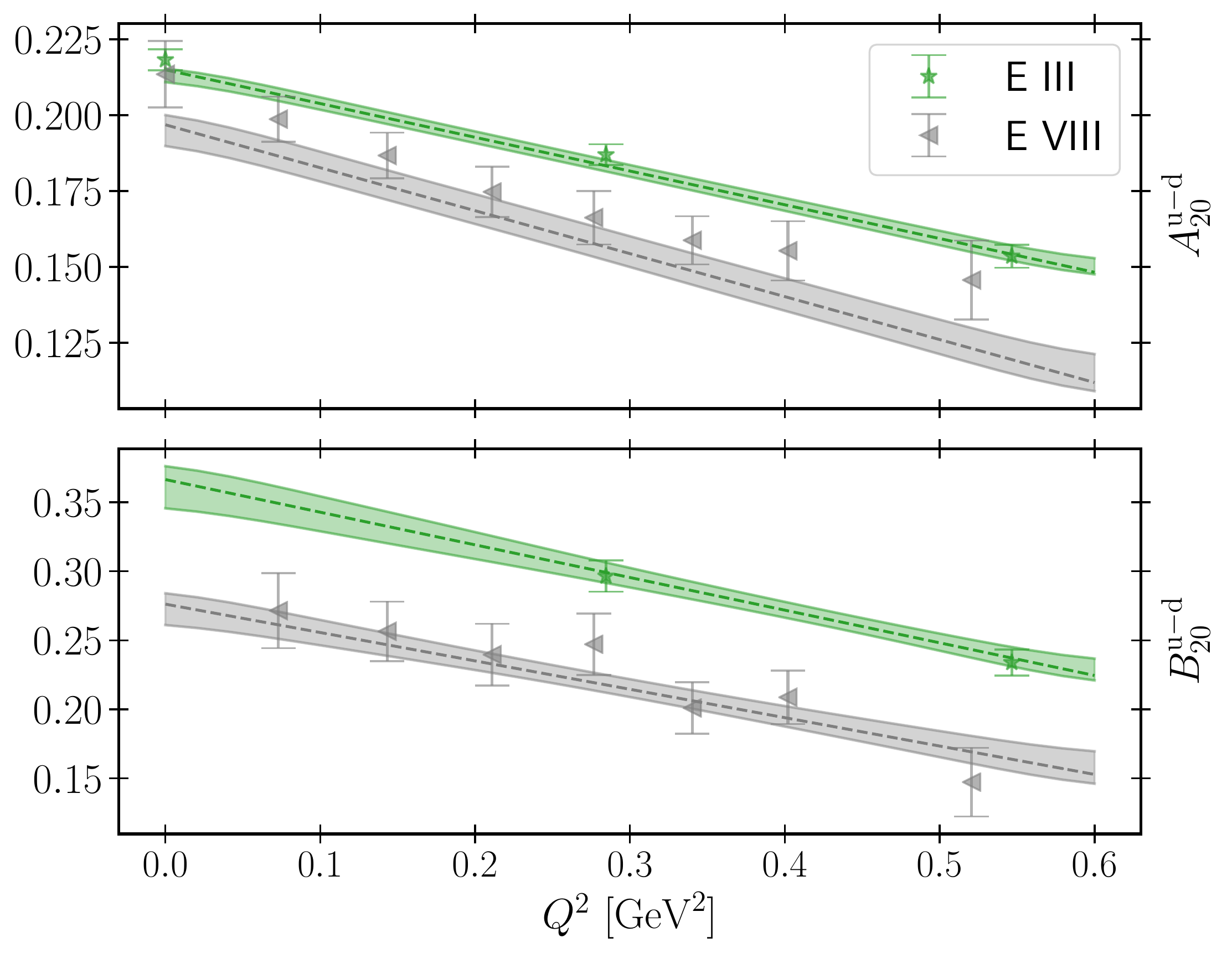}
  \caption{Chiral fit A  versus $Q^2$ for two distinct pion masses:
  	$m_\pi=422\,\text{MeV}$ (green) and $150\,\text{MeV}$ (grey).
  	The corresponding data points (ensemble III and VIII) are shown as well.
  \label{fig:jpp}
  }
\end{figure}

Since it is not clear up to what values of $-\mathsf{t}$ and $m_\pi$ BChPT
is applicable,
we perform fits to all ensembles (set A) as well as fits using only
ensembles
with $m_\pi \le 300 \,\mathrm{MeV}$ (set B). 
  In \Fig{fig:j} we show the resulting fits for $\mathsf{t}=0$,
  where only in the case of $A_{20}^{u-d}$ we can directly compare
  to data points.
For set A the fit parameters have smaller statistical errors.
For set B we see that
$A_{20}^{u-d}$ increases with $m_\pi  \rightarrow  m_{\pi}^{\mathrm{phy}}$. For
both sets we obtain values for $\chi_{\mathrm{dof}}^2$ of about 0.75, hence
we cannot use the $\chi_{\mathrm{dof}}^2$ value to discriminate between the
fit ranges. Instead, one may interpret the difference between fits A and B
as a systematic uncertainty of the parameters. In Fig.~\ref{fig:jpp} we show
  our fit for set A as a function of $Q^2$ at two fixed values
  of the pion masses ($m_\pi=422\,\text{MeV}$
  and $150\,\text{MeV}$, ensembles III and VIII).
  Obviously, our ansatz for the $Q^2$ and $m_\pi^2$ dependence describes the lattice data well.

Again, we study
the effect of the uncertainties of the renormalization constants using
the strategy described in Appendix~\ref{sec:zerr}. The final results are
collected in \Tab{tab:res_j}, where we also quote the
  total angular momentum $J^{u-d}$. We refrain
  from extrapolating to $Q^2=0$ and $m_{\pi}=m_{\pi}^{\mathrm{phy}}$ in the
  other cases. Instead, in \Tab{tab:res_others} we give the results
  for the form factors where no extrapolation in $Q^2$ is required,
  i.e.\ $A_{20}^{u-d}(0)$, $\widetilde{A}_{20}^{u-d}(0)$ and $A_{T20}^{u-d}(0)$,
  for our
  nearly physical point ensemble VIII. The moment $A_{20}^{u-d}(0)=\langle
  x\rangle_{u-d}$ agrees well with the results of the global fits to ensemble
  sets A and B and also the helicity and transversity
  moments $\widetilde{A}_{20}^{u-d}(0)=\langle x\rangle_{\Delta u -\Delta d}$
  and $A_{T20}^{u-d}(0)=\langle x\rangle_{\delta u -\delta d}$ at the physical
  point ensemble are in agreement with the global data, see the top right
  panel of Fig.~\ref{fig:vaGFF} and the top left panel of Fig.~\ref{fig:tGFF},
  respectively. 

Within the errors, our values
agree with the isovector results of Ref.~\cite{Alexandrou:2017oeh}.
\begin{table}[tb]
  \caption{
    Results for $A_{20}^{u-d}(0,m_\pi)$, $B_{20}^{u-d}(0,m_\pi)$ and $J^{u-d}(m_\pi)$, extrapolated to the physical pion mass $m_\pi^{\mathrm{phy}}$ using the ensemble sets A and B (see the text). The first error is statistical, the second error is due to the uncertainty of the renormalization constants.
    \label{tab:res_j}}
  \begin{center}
    \begin{ruledtabular}
      \begin{tabular}{c@{\qquad}cc}
        Ensemble selection        & A                       &        B \\
        \hline
        $A_{20}^{u-d}(0, m_{\pi}^{\mathrm{phy}})$  & $0.195 \,(06) \, (03)$  & $0.210\,(08)\,(04)$\\
        $B_{20}^{u-d}(0, m_{\pi}^{\mathrm{phy}})$  & $0.271 \,(13) \, (03)$  & $0.287\,(28)\,(04)$\\
        $J^{u-d}(m_{\pi}^{\mathrm{phy}})$       & $0.233 \,(07) \, (03)$  & $0.248\,(14)\,(04)$ \\
      \end{tabular}
    \end{ruledtabular}
  \end{center}
  \caption{Results for $A_{20}^{u-d}$, $\widetilde{A}_{20}^{u-d}$ and $A_{T20}^{u-d}$
    at the nearly physical pion mass $m_\pi=150\,\text{MeV}$ (ensemble VIII). The first error is statistical,
    the second error is due to the uncertainty of the renormalization constants.}
    \label{tab:res_others}
  \begin{center}
    \begin{ruledtabular}
      \begin{tabular}{c@{\qquad}c}
        Ensemble VIII    & Value   \\
        \hline
        $A_{20}^{u-d}(0,m_\pi)$  & $0.213 \,(11) \, (04)$ \\        
        $\widetilde{A}_{20}^{u-d}(0,m_\pi)$  & $0.240 \,(07) \, (03)$ \\
        $A_{T20}^{u-d}(0,m_\pi)$  & $0.266 \,(08) \, (04)$
      \end{tabular}
    \end{ruledtabular}
  \end{center}
\end{table}

\section{Nucleon tomography}
\label{sec:nucleon_tomography}
We use our lattice results for the vector GFFs $A_{20}(\mathsf{t})$, $B_{20}(\mathsf{t})$ and the linear combination $\overline{B}_{T20}(\mathsf{t})$
[cf.~Eq.~(\ref{eq:Bbar})] to investigate the transverse spin density of the nucleon. To this end, we transform these GFFs to the impact parameter space $G(\mathsf{t}) \rightarrow  G(\boldsymbol{b}_{\perp}^2)$ with
\begin{align}
	\label{eq:gffb}
    G\left(\boldsymbol{b}_{\perp}^2\right)=\int\frac{\mathrm{d}^2\boldsymbol{\Delta_{\perp}}}{(2\pi)^2}\;
    e^{-i \boldsymbol{b_{\perp}}\cdot \boldsymbol{\Delta_\perp}}\: G\left(\mathsf{t}=-\boldsymbol{\Delta}^2_{\perp}\right)\,,
\end{align}
where we use the $p$-pole ansatz~\cite{Diehl:2005jf,Gockeler2007}
\begin{align}
\label{eq:dipol}
 G(\mathsf{t}) = \frac{G_0}{\left(1 - \mathsf{t}/m_p^2 \right)^p}
\end{align}
for the interpolation of our lattice results.
The impact parameter $\boldsymbol{{b_\perp}}$ is defined in the transverse $x$-$y$ plane.
It measures the transverse distance from the ``center of momentum''
\begin{align}
\boldsymbol{R}_\perp = \sum \limits_i \boldsymbol{r}_{i\,\perp} x_i\,,
\quad \sum \limits_i x_i = 1\,,
\end{align}
where $x_i$ is the momentum fraction of the $i$th parton
\cite{Diehl:2005jf,PhysRevD.15.1141}.
We define
\begin{align}
 \boldsymbol{b}_\perp \coloneqq ( b_x, \, b_y) \, , \quad \quad
  b_{\perp} \coloneqq  \sqrt{ \boldsymbol{b}_\perp^2} \, .
\end{align}
To compute the transverse spin density, we also have to evaluate the derivative of
$G(b_\perp^2 )$ with respect to $b_\perp^2$,
\begin{align}
G^{\prime}(b_\perp^2 )  \coloneqq  \frac{\partial}{\partial \, b_\perp^2 } \, G(b_\perp^2).
\end{align}
The Fourier transform (\ref{eq:gffb}) of the $p$-pole ansatz (\ref{eq:dipol})
can be expressed in terms of the modified Bessel functions $K_\nu$~\cite{Diehl:2005jf},
\begin{align}
	\label{eq:gffb0}
    G( b_{\perp}^2  ) &= \frac{
    G_0 \, m_p^2 \;   (b_{\perp}m_p)^{p-1}  \, K_{p-1}( b_{\perp} m_p)
    }{
	    2^p \, \pi\, \Gamma(p)
    }\,.
\end{align}
The transverse spin density $\rho^q(x,\boldsymbol{{b_\perp}},\boldsymbol{s_\perp},\boldsymbol{S_\perp})$
describes the probability to find a quark with longitudinal momentum fraction $x$,
flavor $q$ and transverse spin $\boldsymbol{s}_\perp$ at a distance $\boldsymbol{b}_{\perp}$
from the center of momentum of the nucleon with transverse spin $\boldsymbol{S}_\perp$.
The explicit definition in terms of GPDs is given in Eq.\,(8) of Ref.~\cite{Diehl:2005jf}.
Here we consider the two transverse spin combinations,
\begin{subequations}
\label{eq:ss}
\begin{align}
 \label{eq:s10_S00}
   \boldsymbol{s}_\perp = (1 \,, 0) \quad &\text{and} \quad \boldsymbol{S}_\perp = (0 \,, 0)\,,\\
   \boldsymbol{s}_\perp = (0 \,, 0) \quad &\text{and} \quad \boldsymbol{S}_\perp = (1 \,, 0)\,,
\end{align}
\end{subequations}
where the first line describes a transversely polarized quark in an unpolarized nucleon and
the second an unpolarized quark in a transversely polarized nucleon.
In terms of GFFs the first moment of
$\rho^q(x, \boldsymbol{{b_\perp}}, \boldsymbol{s_\perp},\boldsymbol{S_\perp})$
for these spin combinations reads
\begin{align}
\langle\rho\rangle^q(\boldsymbol{b_\perp},\boldsymbol{s_\perp},\boldsymbol{S_\perp})  &=
\int_{-1}^{1}\!\mathrm{d}x\,x \, \rho^q(x, \boldsymbol{{b_\perp}}, \boldsymbol{s_\perp},\boldsymbol{S_\perp})
\nonumber \\
=\frac{1}{2}A_{20}^{q}(b_\perp^2)
&-\frac{ \epsilon^{ij} \,  b_\perp^j }{2m_N}  \left( s_\perp^i \overline{B}_{T20}^{q\,\prime}(b_\perp^2)  +  S_\perp^i B_{20}^{q\,\prime}(b_\perp^2)  \right) \,.
\label{eq:density_simpl}
\end{align}
For arbitrary spins $\boldsymbol{S}_\perp$ and $\boldsymbol{s}_\perp$
\Eq{eq:density_simpl} will contain additional terms and we refer the reader to Refs.~\cite{Diehl:2005jf,Gockeler2007}.

\begin{figure}[tb]
	\centering
   \includegraphics[width=0.5\textwidth]{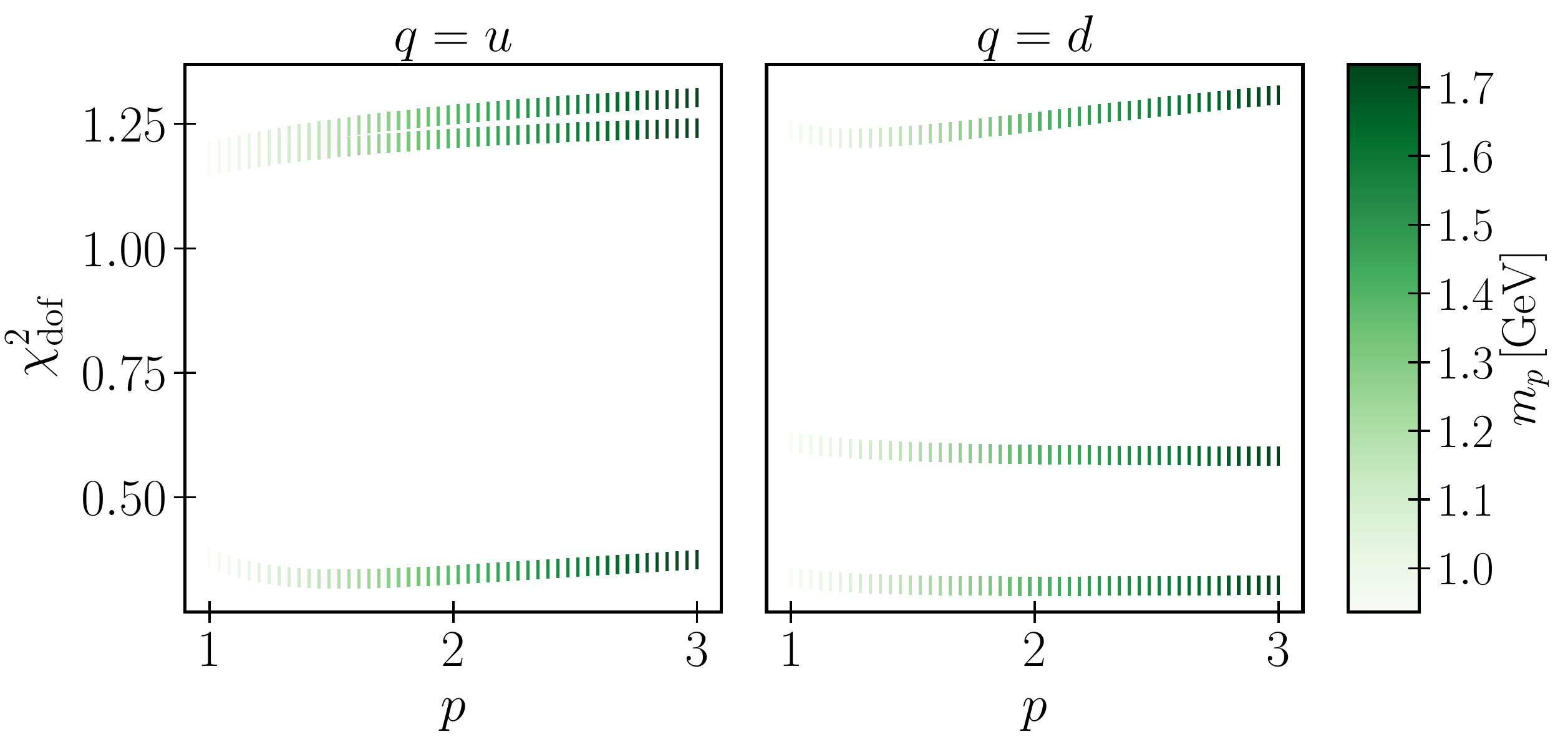}
   \caption{
   The pole mass $m_p$ and $\chi^2_{\mathrm{dof}}$
   as a function of the fixed parameter $p$ for ensemble VI.
   The colored lines correspond to fits to
   $A_{20}^q$,
   $B_{20}^q$ and
   $\overline{B}_{T20}^q$ from top to bottom
   and flavor $q$ from left to right.
\label{fig:dipol_scan}
}
\end{figure}

\begin{figure}[tb]
	\centering
   \includegraphics[width=0.45\textwidth]{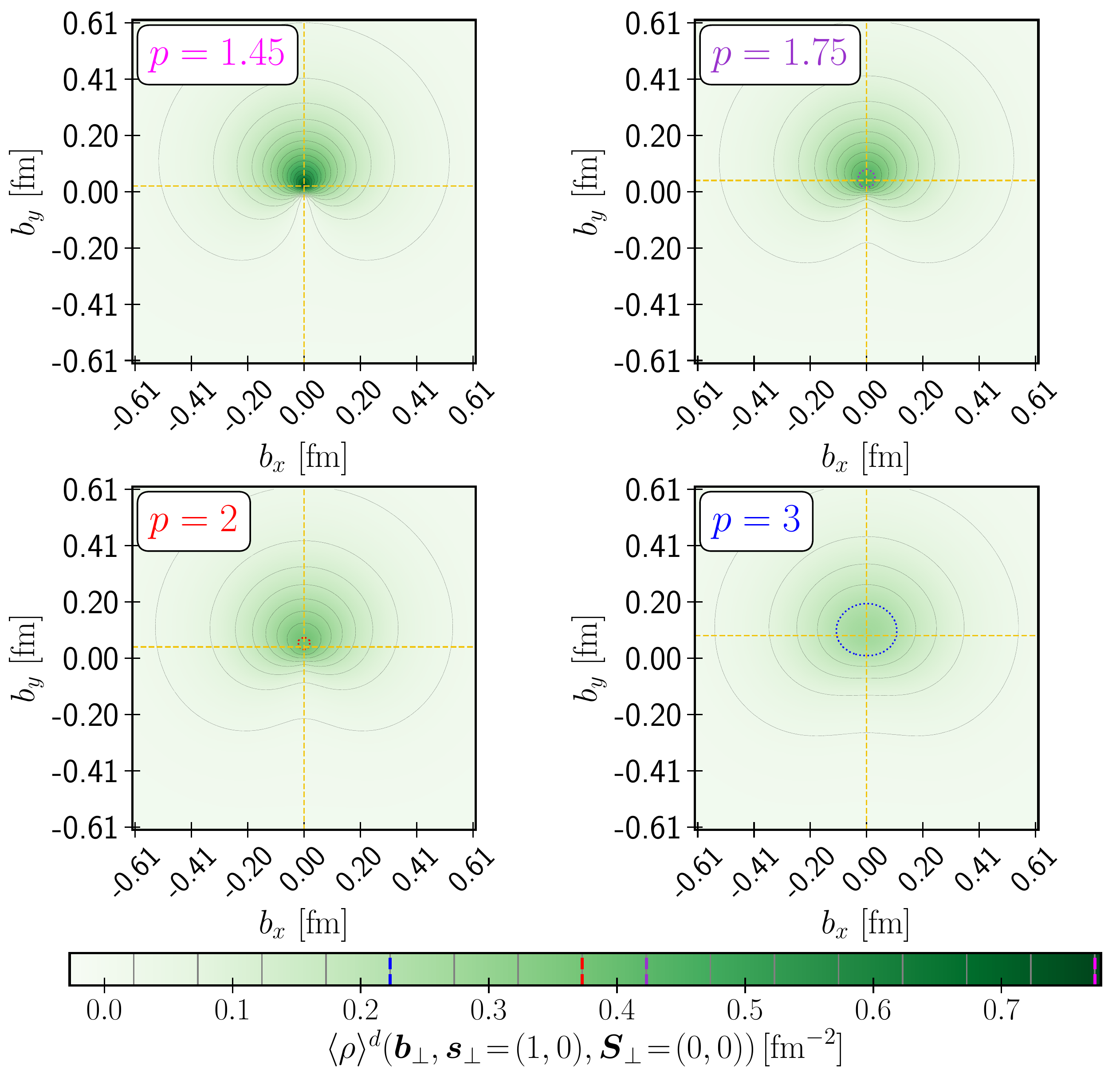}
   \caption{
   The $p$-dependence of the transverse spin density for a
   transversely polarized $d$-quark in an unpolarized nucleon.
   The yellow cross indicates the maximum of the density.
   The black contour lines are drawn equidistantly with a
     difference of 0.05.\label{fig:dens_scan}
}
\end{figure}

We fit the GFFs for ensemble VI to the $p$-pole ansatz \Eq{eq:dipol}.
Due to the limited number of data points at our disposal, where we restricted
ourselves to the kinematic range $-\mathsf{t} \le 0.6 \, \mathrm{GeV}^2$,
we find it impossible to simultaneously determine all three fit parameters, $p$, $m_p$ and $G_0$. In particular the exponent $p$ is strongly correlated with the pole mass $m_p$. This is demonstrated in \Fig{fig:dipol_scan}: An increase of $p$ results in a larger value of $m_p$, whereas $\chi^2_{\mathrm{dof}}$ does not significantly change. Therefore, we cannot constrain $p$.

This arbitrariness means it is difficult to obtain reliable, parametrization
independent results for the moment $\langle\rho\rangle^q(\boldsymbol{b_\perp},\boldsymbol{s_\perp},\boldsymbol{S_\perp})$ as a function of $\boldsymbol{b_\perp}$. This distribution has been studied in the past (see, e.g.,~\cite{Gockeler2007}),
but we find that its shape strongly depends on the value of $p$.
In \Fig{fig:dens_scan} we show
$\langle\rho\rangle^q(\boldsymbol{b_\perp},\boldsymbol{s_\perp},\boldsymbol{S_\perp})$
for $\boldsymbol{s}_\perp = (1,0)$ and $\boldsymbol{S}_\perp = (0, 0)$ for four distinct values of $p$ ranging from
1.45 up to 3.0.
We see that with increasing $p$ the density becomes less localized in
the impact parameter plane and the maximum of the density is shifted away from the center.
This also holds for other spin combinations.

\begin{table}[t]
  \caption{The half $\boldsymbol{b_\perp}$-integrated moments
      for $p=2$, also shown in Fig.~\ref{fig:rho_pm}. The errors are
      statistical. The systematic error of the $p$-dependence is about
      0.02.
    \label{tab:res_rho_pm}}
  \begin{center}
    \begin{ruledtabular}
      \begin{tabular}{ccc}
& $\boldsymbol{s}_\perp = (1 \,, 0)$  &   $\boldsymbol{s}_\perp = (0 \,, 0) $   \\
& $\boldsymbol{S}_\perp = (0 \,, 0)$  &   $\boldsymbol{S}_\perp = (1 \,, 0) $   \\
 \hline
    $\langle\rho\rangle^{u}_{-}$         &    0.312 (26) &  0.403 (12) \\
    $\langle\rho\rangle^{u}_{+}$         &    0.688 (26) &  0.597 (12) \\
    \hline
    $\langle\rho\rangle^{d}_{-}$         &    0.262 (27) &  0.666 (17) \\
    $\langle\rho\rangle^{d}_{+}$         &    0.738 (27) &  0.334 (17)
      \end{tabular}
    \end{ruledtabular}
  \end{center}
\end{table}

\begin{figure}[tb]
	\centering
   \includegraphics[width=0.45\textwidth]{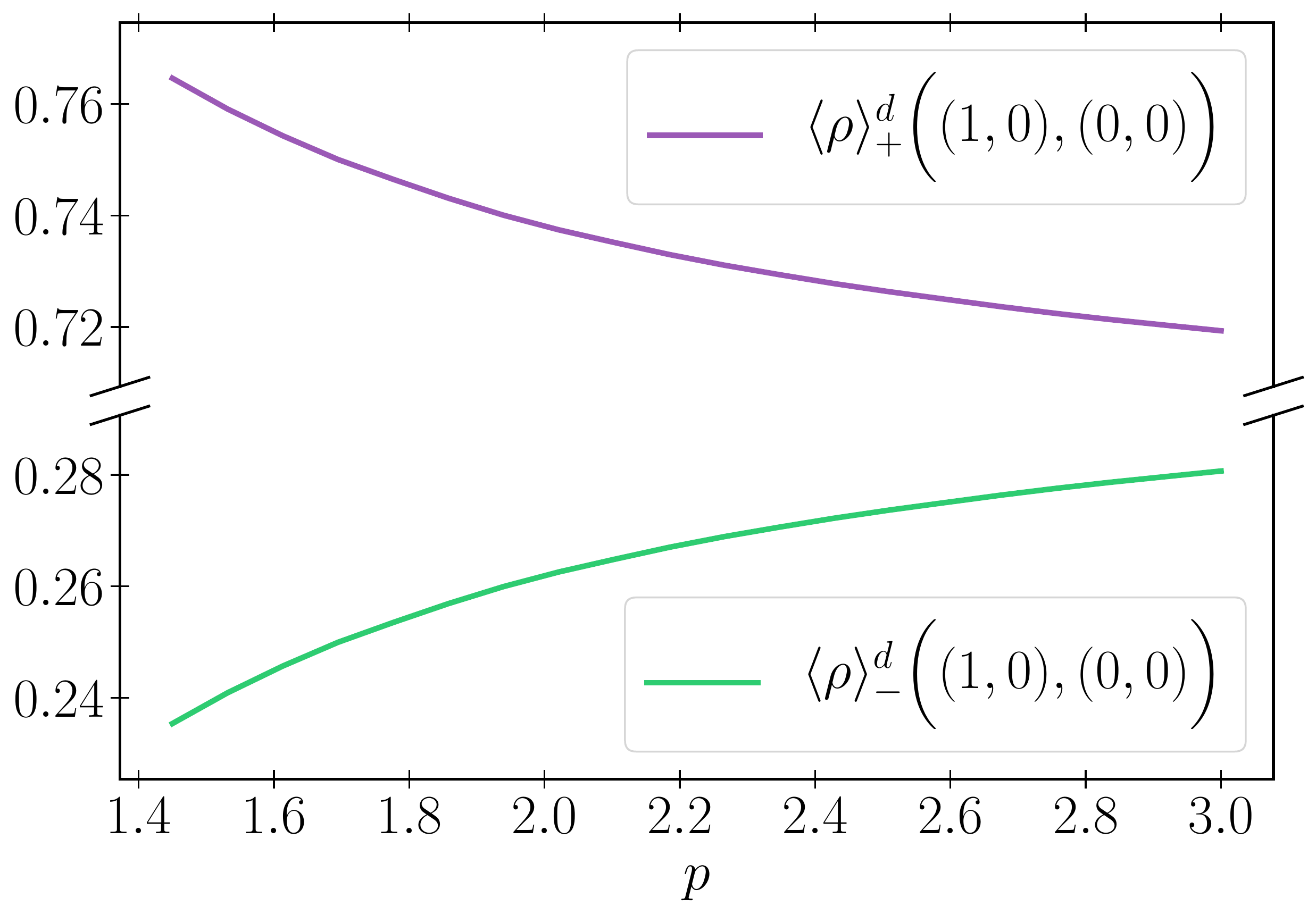}
   \caption{Dependence of $\langle\rho\rangle^d_{+}\left(\boldsymbol{s_\perp},\boldsymbol{S_\perp}\right)$ and
	$\langle\rho\rangle^d_{-}\left(\boldsymbol{s_\perp},\boldsymbol{S_\perp}\right)$
     on the power $p$ of the pole ansatz. The combination of transverse spins is $\boldsymbol{s_\perp}=(1,0)$ and $\boldsymbol{S_\perp}=(0,0)$. The errors
       are statistical only. The systematics due to the uncertainty of the power
       $p$ amount to about 0.02.
	\label{fig:densp_scan}
}
\end{figure}

\begin{figure}[tb]
	\centering
   \includegraphics[width=0.45\textwidth]{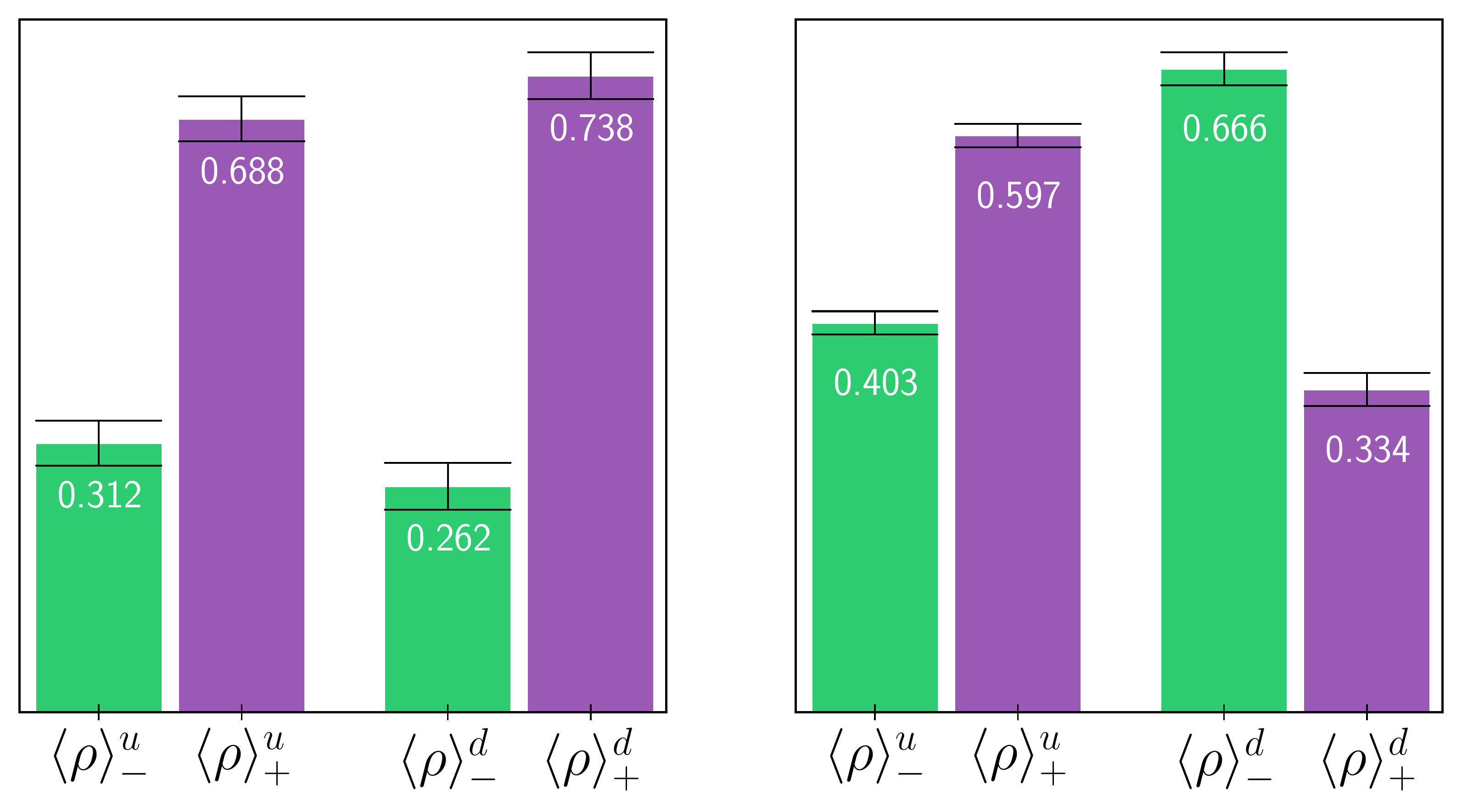}
   \caption{Probability (weighted with $x$) to find a $u$- or $d$-quark in the upper/ lower part ($b_y\lessgtr 0$) of the impact parameter space; left for a transversely polarized quark in an unpolarized nucleon; right for an unpolarized quark in a transversely polarized nucleon.
\label{fig:rho_pm}
}
\end{figure}

We discovered that some integrated quantities have a much
milder $p$-dependence, namely the half $\boldsymbol{b_\perp}$-integrated moments
\begin{subequations}
\label{eq:rho_pm}
	\begin{align}
	\langle\rho\rangle^q_{+}(\boldsymbol{s_\perp},\boldsymbol{S_\perp})  &=
	\frac{1}{Z_{\rho}}
	\int \limits_{-\infty}^{+\infty}\!
	\mathrm{d}b_x \! \!
	\int \limits_{0}^{+\infty}\!
	\mathrm{d}b_y \,
	\langle\rho\rangle^q(\boldsymbol{b_\perp},\boldsymbol{s_\perp},\boldsymbol{S_\perp}) \, ,
	\\
	\langle\rho\rangle^q_{-}(\boldsymbol{s_\perp},\boldsymbol{S_\perp})  &=
	\frac{1}{Z_{\rho}}
	\int \limits_{-\infty}^{+\infty}\!
	\mathrm{d}b_x  \!\!
	\int \limits_{-\infty}^{0}\!
	\mathrm{d}b_y \,
	\langle\rho\rangle^q(\boldsymbol{b_\perp},\boldsymbol{s_\perp},\boldsymbol{S_\perp}) \,,
\end{align}
\end{subequations}
with the normalization factor
\begin{align}
	{Z_{\rho}} &=
	\int \limits_{-\infty}^{+\infty}\!
	\mathrm{d}b_x \,
	\int \limits_{-\infty}^{+\infty}\!
	\mathrm{d}b_y \, \langle\rho\rangle^q(\boldsymbol{b_\perp},\boldsymbol{s_\perp},\boldsymbol{S_\perp}) \,.
\end{align}
The integrated moment $\langle\rho\rangle^q_{+}(\boldsymbol{s_\perp},\boldsymbol{S_\perp})$
is the probability, weighted with the longitudinal momentum fraction $x$, to find a quark with flavor $q$ in the upper part ($b_y\ge 0$) of the impact parameter space
and $\langle\rho\rangle^q_{-}(\boldsymbol{s_\perp},\boldsymbol{S_\perp})$ is the
$x$-weighted probability to find a quark with flavor $q$ in the lower part ($b_y\le 0$).
These integrated moments are a measure for the asymmetry of the
transverse spin density.
They depend much less on the value of $p$ than
$\langle\rho\rangle^q(\boldsymbol{b_\perp},\boldsymbol{s_\perp},\boldsymbol{S_\perp})$
does.
This is demonstrated in \Fig{fig:densp_scan},
where $\langle\rho\rangle^d_{+}$
and $\langle\rho\rangle^d_{-}$ are shown as functions of $p$ for
the transverse spin combination in
\Eq{eq:s10_S00}.
Doubling $p$, both integrated moments change
by only 5\% and 15\%, respectively.
We find this mild $p$-dependence for all considered transverse spin and flavor
combinations and consider these integrated moments as the better
candidates for reliable lattice estimates.
Our results for $\langle\rho\rangle^q_{\pm}$ for up and down quark
for our two transverse spin combinations [\Eq{eq:ss}]
are shown in \Fig{fig:rho_pm}. The errors shown
  are statistical only.
  The figure corresponds to the power $p=2$, and one may add
  systematic errors of about 0.02 due to the $p$-dependence;
  see \Fig{fig:densp_scan}. The numerical values are listed in
  \Tab{tab:res_rho_pm}.

We see the probability of a transversely polarized $u$- or $d$-quark in an
unpolarized nucleon is higher ($\sim 70\%$)
in the $b_y>0$ part of the impact parameter space than in the
$b_y<0$ part ($\sim30\%$).
For a transversely polarized nucleon however the probabilities of an
unpolarized $u$- or $d$-quark differ:
The unpolarized $d$-quark is more likely in the
$b_y<0$ part ($67\%$),
while a $u$-quark is more likely in the $b_y>0$ part (60\%)
of the impact parameter space.


\section{Summary}
\label{sec:summary}
We have calculated all quark GFFs, corresponding to operators
with one derivative, of the nucleon GPDs at leading twist-2.
Our lattice calculation includes the dominating connected contributions and
neglects contributions from disconnected diagrams.
The available gauge ensembles cover a wide range of
quark masses and volumes.
However, the three available
lattice spacings only vary from 0.081\,fm down to 0.060\,fm.
Within errors, all GFFs show a mild dependence on the quark mass,
lattice spacing and volume.

We have compared two different fitting strategies for the GFFs and found that
the direct fit method appears to be more reliable.
With this method the number of fit parameters is reduced to the relevant
degrees of freedom. We recommend to use this method in future studies.
The final results for the GFFs are shown in Figs.~\ref{fig:vaGFF}
and~\ref{fig:tGFF}.

We have also studied the total angular momentum and the
transverse spin density of quarks in the nucleon.
Both quantities can be extracted from fits to our GFF data.
For the total angular momentum we obtain a similar estimate in the isovector
case as ETMC in Ref.~\cite{Alexandrou:2017oeh}.
Contributions from disconnected diagrams are not included in our lattice calculation. From Ref.~\cite{Alexandrou:2017oeh} we know that these
are small. Nevertheless, in the isoscalar case they should definitely be taken
into account.
For the second moment of the transverse spin density we have found that
its distribution in impact parameter space strongly depends
on the $\mathsf{t}$-dependence of the GFF data.
The shape of the distribution depends on the value of $p$ that is
used within a $p$-pole ansatz.
High precision data at small and large values of $-\mathsf{t}$ would
be required to eliminate this ambiguity.
For integrated moments this situation improves.
In \Fig{fig:rho_pm} we provide lattice estimates for the
$x$-weighted probabilities of a transversely polarized (unpolarized)
light quark in the upper or lower part of the impact parameter space,
within an unpolarized (transversely polarized) nucleon.
Contributions from higher moments are not yet available but
constitute an interesting object for future study.

\begin{acknowledgments}
  The ensembles were generated by RQCD and QCDSF primarily on the QPACE computer~\cite{Baier:2009yq,Nakamura:2011cd}, which was built as part of the
DFG (SFB/TRR 55) project. The authors gratefully acknowledge the Gauss Centre
for Supercomputing e.V.\ (\href{http://www.gauss-centre.eu}{\url{www.gauss-centre.eu}})
for granting computer time on SuperMUC at the Leibniz Supercomputing Centre
(LRZ, \href{http://www.lrz.de}{\url{www.lrz.de}}) for this project.
The BQCD~\cite{Nakamura:2010qh} and CHROMA~\cite{Edwards:2004sx} software
packages were used, along with the locally deflated domain decomposition solver
implementation of openQCD~\cite{Luscher:2012av,openQCD}.
Part of the analysis was also performed on the iDataCool cluster in Regensburg.
Support was provided by the DFG (SFB/TRR 55).
ASt acknowledges support by the BMBF under Grant
No.\ 05P15SJFAA (FAIR-APPA-SPARC) and by the DFG Research
Training Group GRK1523.
We thank Benjamin Gl\"a\ss{}le for software support.
\end{acknowledgments}

\appendix
\section{Operator multiplets for the tensor GFFs}
\label{app:tensor_olc}
In this study we use 16 linear combinations of operators for the tensor GFFs. The first eight from the $h_{1,a}$ multiplet read
\begin{align*}
\mathcal{O}_{1}^{h_{1,a}}   &= \sqrt{\frac{2}{3}}( \mathcal{O}_{132}^{\intercal} + \frac{1}{2}\mathcal{O}_{123}^{\intercal} +\frac{1}{2}\mathcal{O}_{231}^{\intercal}), \\
\mathcal{O}_{2}^{h_{1,a}}   &= \sqrt{\frac{2}{3}}( \mathcal{O}_{142}^{\intercal} + \frac{1}{2}\mathcal{O}_{124}^{\intercal} +\frac{1}{2}\mathcal{O}_{241}^{\intercal}), \\
\mathcal{O}_{3}^{h_{1,a}}   &= \sqrt{\frac{2}{3}}( \mathcal{O}_{143}^{\intercal} + \frac{1}{2}\mathcal{O}_{134}^{\intercal} +\frac{1}{2}\mathcal{O}_{341}^{\intercal}), \\
\mathcal{O}_{4}^{h_{1,a}}   &= \sqrt{\frac{2}{3}}( \mathcal{O}_{243}^{\intercal} + \frac{1}{2}\mathcal{O}_{234}^{\intercal} +\frac{1}{2}\mathcal{O}_{342}^{\intercal}), \\
\mathcal{O}_{5}^{h_{1,a}}   &= \sqrt{2} \mathcal{O}_{2 \{13\}}^{\intercal}, \\
\mathcal{O}_{6}^{h_{1,a}}   &= \sqrt{2} \mathcal{O}_{2 \{14\}}^{\intercal}, \\
\mathcal{O}_{7}^{h_{1,a}}   &= \sqrt{2} \mathcal{O}_{3 \{14\}}^{\intercal}, \\
\mathcal{O}_{8}^{h_{1,a}}   &= \sqrt{2} \mathcal{O}_{3 \{24\}}^{\intercal}\,. \\
\end{align*}
The remaining eight make up the $h_{1,b}$ multiplet and read
\begin{align*}
\mathcal{O}_{9}^{h_{1,b}}   &= \sqrt{\frac{1}{2}}( \mathcal{O}_{122}^{\intercal} - \mathcal{O}_{133}^{\intercal}), \\
\mathcal{O}_{10}^{h_{1,b}}  &= \sqrt{\frac{1}{2}}( \mathcal{O}_{211}^{\intercal} - \mathcal{O}_{233}^{\intercal}), \\
\mathcal{O}_{11}^{h_{1,b}}  &= \sqrt{\frac{1}{2}}( \mathcal{O}_{311}^{\intercal} - \mathcal{O}_{322}^{\intercal}), \\
\mathcal{O}_{12}^{h_{1,b}}  &= \sqrt{\frac{1}{2}}( \mathcal{O}_{411}^{\intercal} - \mathcal{O}_{422}^{\intercal}), \\
\mathcal{O}_{13}^{h_{1,b}}  &= \sqrt{\frac{1}{6}}(\mathcal{O}_{122}^{\intercal} + \mathcal{O}_{133}^{\intercal} - 2\mathcal{O}_{144}^{\intercal}), \\
\mathcal{O}_{14}^{h_{1,b}}  &= \sqrt{\frac{1}{6}}(\mathcal{O}_{211}^{\intercal} + \mathcal{O}_{233}^{\intercal} - 2\mathcal{O}_{244}^{\intercal}), \\
\mathcal{O}_{15}^{h_{1,b}}  &= \sqrt{\frac{1}{6}}(\mathcal{O}_{311}^{\intercal} + \mathcal{O}_{322}^{\intercal} - 2\mathcal{O}_{344}^{\intercal}), \\
\mathcal{O}_{16}^{h_{1,b}}  &= \sqrt{\frac{1}{6}}(\mathcal{O}_{411}^{\intercal} + \mathcal{O}_{422}^{\intercal} - 2\mathcal{O}_{433}^{\intercal}).
\end{align*}

\section{Renormalization procedure}
\label{app:renormproc}
The renormalization factors are products of
perturbative and nonperturbative parts:
\begin{equation}
 Z^{\MS}_{\mathcal{O}}
\coloneqq Z_{\mathcal{O},\RI}^\MS Z_{\mathcal{O},\mathrm{bare}}^\RI\,.
\end{equation}
The nonperturbative factor $Z^{\RI}_{\mathcal{O},\mathrm{bare}}$ translates
the bare lattice data to the regularization scheme independent momentum
subtraction ($\RIMOM$) scheme~\cite{Martinelli:1994ty,Chetyrkin:1999pq},
while the perturbative factor $Z_{\mathcal{O},\RI}^\MS$ matches from
the $\RIMOM$ to the $\MS$ scheme.
This is calculated in continuum perturbation theory
and is known for our operator multiplets to three-loop accuracy~\cite{Gracey:2003mr}.

\subsection{Nonperturbative renormalization}
\label{sec:RIMOM}
The nonperturbative renormalization factors
$Z^{\RI}_{\mathcal{O},\mathrm{bare}}$ are extracted as follows.
In a first step we gauge-fix a subset\footnote{About ten
well-decorrelated configurations are often sufficient.} of our gauge
configurations to Landau gauge and calculate (in momentum space)
the quark propagator
\begin{equation}
   S_{\alpha\beta}(a,p) = \frac{a^{8}}{V}\sum_{xy}
      e^{-ip\cdot(x-y)}\left\langle q_\alpha(x)\bar{q}_\beta(y)\right\rangle
\end{equation}
(color indices are suppressed) and the three-point functions
\begin{equation}
  G^{{j,\mu}}_{\alpha\beta}(a,p) =
       \frac{a^{12}}{V} \sum_{xyz}  e^{-ip\cdot(x-y)} \left\langle
   q_\alpha(x)\,\mathcal{J}^j_{\mu}(z)\, \bar{q}_\beta(y)\right\rangle
\end{equation}
with $\mathcal{J}^j_{\mu}(z) \coloneqq \bar{q}(z)\,\mathsf{\Gamma}^j
\overleftrightarrow{\nabla}_{\!\mu} q(z)=\bar{q}(z)\mathcal{O}^j_{\mu}(z)q(z)$.
$\mathsf{\Gamma}^j$
denotes one of the sixteen possible products of
Euclidean gamma matrices, $\gamma^{n_1}_1\cdots\gamma_4^{n_4}$
($n_\mu\in\{0,1\}$), and the covariant lattice derivative acts on the
respective left or right quark propagators resulting from the
integration over the quark fields.

Next the vertex function $\Gamma_\mathcal{O}$ is constructed for each
operator $\mathcal{O}(0)$ by combining the appropriate $G^{j,\mu}$s
and amputating the fermion legs,
\begin{equation}
  \Gamma_\mathcal{O}(a,p) \coloneqq
   S^{-1}(a,p)G_{\mathcal{O}}(a,p)S^{-1}(a,p)\,.
\end{equation}
The renormalized vertex reads
\begin{equation}
 \Gamma^R_\mathcal{O}(p,\mu^2) =
      \frac{Z^{\RI}_{\mathcal{O},\mathrm{bare}}(a,\mu^2)}{Z_q(a,\mu^2)}
    \Gamma_\mathcal {O}(a,p)\,,
\end{equation}
where the $\RIMOM$ renormalization condition
\begin{equation}
  \left.\frac{1}{12}\Tr\left(\Gamma^R_\mathcal{O}
  \left[\Gamma^{(0)}_\mathcal{O}\right]^{-1}\right) \stackrel{!}{=} 1
  \quad\right\vert_{p^2=\mu^2}
\label{eq:RIMOM_condition}
\end{equation}
is imposed in the chiral limit. The quark wave
function renormalization factor is given by
\begin{equation}
 Z_q(a,\mu^2) = \left.\frac{-i\Tr\left( \gamma_\lambda \bar{p}_\lambda \,
 S^{-1}(a,p)\right)}{12 \bar{p}^2 }\right\vert_{p^2=\mu^2}
\label{eq:quarkwave}
\end{equation}
after extrapolation to the massless limit.
In Eq.~\eqref{eq:quarkwave} we employ the lattice tree-level expression for
the massless quark propagator; i.e., we set $a\bar{p}_\lambda \coloneqq
\sin(ap_\lambda)$. Similarly we use the lattice tree-level expression for
the Born term $\Gamma^{(0)}_{\mathcal{O}}$ to reduce lattice discretization
effects. For the example of the operator $\mathcal{O}^{\mu\nu}_{V,q}$ this reads
\begin{equation}
  \Gamma^{(0)}_\mathcal{O}(p) = i\left(\gamma_\mu\bar{p}_\nu +
\gamma_\nu\bar{p}_\mu\right)\,.
\end{equation}

\subsection{Propagation of renormalization constant errors}
\label{sec:zerr}
Our estimates for the renormalization factors carry an uncertainty which
has to be propagated into the GFFs. We do this in a very naive but
conservative way by carrying out the whole analysis both using the
central values of the renormalization factors and adding the error
of these factors to their central values. The difference between these
two sets of results is then due to the uncertainty of the renormalization.
This procedure is applied to all ensembles and to all the available
virtualities $Q^2$. We find that the relative error is almost independent
of $Q^2$ and the considered ensemble. Hence, for each GFF we decided to take
the largest value of this uncertainty as an estimator of the error. These
relative uncertainties are shown in \Tab{tab:pe}.
\bibliographystyle{apsrev4-1}
\bibliography{references}

\end{document}